\documentclass[12pt,preprint]{aastex}

\usepackage{lscape}

\usepackage{graphicx}
\usepackage{CJK}

\shorttitle{Class I methanol maser in GLIMPSE and BGPS}
\shortauthors{Xi Chen et al.}

\begin{document}


\title {A 95 GHz Class I Methanol Maser Survey Toward A Sample of GLIMPSE Point Sources Associated with BGPS Clumps}
\author
 {Xi ~Chen\altaffilmark{1, 2}, Simon P. Ellingsen\altaffilmark{3}, Jin-Hua He\altaffilmark{4},
Ye Xu\altaffilmark{5}, Cong-Gui Gan\altaffilmark{1, 2}, Zhi-Qiang
~Shen\altaffilmark{1, 2}, Tao An\altaffilmark{1,2}, Yan
Sun\altaffilmark{5}, Bing-Gang Ju \altaffilmark{5}}

\altaffiltext{1} {Key Laboratory for Research in Galaxies and
Cosmology, Shanghai Astronomical Observatory, Chinese Academy of
Sciences, Shanghai 200030, China; chenxi@shao.ac.cn}

\altaffiltext{2} {Key Laboratory of Radio Astronomy, Chinese Academy
of Sciences, China}

\altaffiltext{3} {School of Mathematics and Physics, University of
Tasmania, Hobart, Tasmania, Australia}

\altaffiltext{4} {National Astronomical Observatories/Yunnan
Observatory, Chinese Academy of Sciences, Kunming 650011, China}

\altaffiltext{5} {Purple Mountain Observatory, Chinese Academy of
Sciences, Nanjing 210008, China }

\label{firstpage}


\begin{abstract}

We report a survey with the Purple Mountain Observatory (PMO) 13.7-m
radio telescope for class I methanol masers from the 95 GHz ($8_{0}$
-- $7_{1}$A$^{+}$) transition.  The 214 target sources were selected
by combining information from both the \emph{Spitzer} GLIMPSE and
1.1 mm BGPS survey catalogs. The observed sources satisfy both the
GLIMPSE mid-IR criteria of [3.6]-[4.5]$>$1.3, [3.6]-[5.8]$>$2.5,
[3.6]-[8.0]$>$2.5 and 8.0 $\mu$m magnitude less than 10, and also
have an associated 1.1 mm BGPS source. Class I methanol maser
emission was detected in 63 sources, corresponding to a detection
rate of 29\% for this survey. For the majority of detections (43),
this is the first identification of a class I methanol maser
associated with these sources. We show that the intensity of the
class I methanol maser emission is not closely related to mid-IR
intensity or the colors of the GLIMPSE point sources, however, it is
closely correlated with properties (mass and beam-averaged column
density) of the BGPS sources. Comparison of measures of star
formation activity for the BGPS sources with and without class I
methanol masers indicate that the sources with class I methanol
masers usually have higher column density and larger flux density
than those without them. Our results predict that the criteria
$log(S_{int})\leq-38.0+1.72log(N_{H_{2}}^{beam})$ and
$log(N_{H_{2}}^{beam})\geq 22.1$, which utilizes both the integrated
flux density (S$_{int}$) and beam-averaged column density
($N_{H_{2}}^{beam}$) of the BGPS sources, are very efficient for
selecting sources likely to have an associated class I methanol
maser.  Our expectation is that searches using these criteria will
detect 90\% of the predicted number of class I methanol masers from
the full BGPS catalog ($\sim$ 1000), and do so with a high detection
efficiency ($\sim$75\%).

\end{abstract}


\section{Introduction}

Methanol masers from a number of transitions are common in active
star forming regions (SFRs) and have been empirically classified
into two categories (class I and class II). Initial studies found
that strong emission from the two classes are preferentially found
towards different star formation regions (Batrla et al. 1988; Menten
1991). Class I methanol masers (e.g. the $7_{0}$ -- $6_{1}$ A$^{+}$
and $8_{0}$ -- $7_{1}$ A$^{+}$ at 44 and 95 GHz respectively) are
typically observed in multiple locations across the star-forming
region spread over an area up to a parsec in extent (e.g. Plambeck
\& Menten 1990; Kurtz et al. 2004; Voronkov et al. 2006; Cyganowski
et al. 2009). In contrast class II methanol masers (e.g. the $5_{1}$
-- $6_{0}$ A$^{+}$ and $2_{0}$ -- $3_{-1}$ E at 6.7 and 12.2 GHz
respectively) are often associated with ultracompact (UC) H{\sc ii}
regions, infrared sources and OH masers and reside close to (within
1$\arcsec$) a high-mass young stellar objects (YSO) (e.g., Caswell
et al. 2010). See M\"{u}ller et al. (2004) for accurate rest
frequencies and other basic data on methanol maser transitions.  The
empirical classification and observational findings were supported
by early theoretical models of methanol masers which suggest that
the pumping mechanism of class I masers is dominated by collisions
with molecular hydrogen, in contrast to class II masers which are
pumped by external far-infrared radiation (Cragg et al. 1992). More
recent modelling has found that in some cases weak class II methanol
masers can be associated with strong class~I masers and vice versa
(e.g. Voronkov et al. 2005), i.e., bright masers of different
classes can not reside in the same volume of gas. High spatial
resolution observations (e.g. Cyganowski et al. 2009) suggest that
where both masers are seen in the same vicinity, while the two types
of masers are not co-spatial on arcsecond scales, they are often
driven by the same young stellar object.

A number of surveys have been performed for class II methanol masers
especially at 6.7 GHz, resulting in the detection of $\sim$ 900
class II maser sources in the Galaxy to date (e.g., the surveys
summarized in the compilation of Pestalozzi et al. 2005 and the
recent searches of Ellingsen 2007, Pandian et al. 2007, Xu et al.
2008, 2009, Green et al. 2009, 2010, 2012 and Caswell et al. 2010,
2011). Class I methanol masers are less well studied than class II
masers, but have recently become the focus of more intense research
(e.g., Sarma \& Momjian 2009, 2011; Fontani et al. 2010; Kalenskii
et al. 2010; Voronkov et al. 2010a, b, 2011; Chambers et al. 2011;
Chen et al. 2011; Fish et al. 2011; Pihlstr\"{o}m et al. 2011).
Early studies of class I masers include only a small number of large
surveys (mainly at 44 and 95 GHz), primarily undertaken with
single-dish telescopes (e.g. Haschick et al. 1990; Slysh et al.
1994; Val'tts et al. 2000; Ellingsen 2005) along with a few smaller
scale interferometric searches (e.g. Kurtz et al 2004, Cyganowski et
al. 2009). Recently, some surveys have been done at other
transitions of class I methanol, e.g. 9.9 GHz by Voronkov et al.
(2010a) and a new class I methanol maser transition at 23.4 GHz has
been discovered by Voronkov et al. (2011). Interferometric
observations show that the class I methanol masers at different
transitions (e.g., 36 GHz and 44 GHz) usually have similar
larger-scale spatial distributions, but are rarely found to produce
a maser at the same site (e.g., Fish et al. 2011). Surveys have
revealed that class I methanol maser (unlike class II masers) are
associated not only with high-mass star formation, but also lower
mass counterparts (Kalenskii et al. 2010).

Recently Chen et al. (2009) demonstrated that a new sample of
massive young stellar object (MYSO) candidates associated with
ongoing outflows (known as extended green objects or EGOs and
identified from the \emph{Spitzer} GLIMPSE survey by Cyganowski et
al. (2008)), provide another productive target for class I maser
searches. On the basis of their statistical analysis Chen et al.
predicted a detection rate of 67\% for class I masers toward EGOs. A
follow-up systemic survey towards a complete EGO sample (192
sources) with the Australia Telescope National Facility (ATNF) Mopra
22-m radio telescope resulted in the detection of 105 new 95 GHz
class I methanol masers (Chen et al. 2011). The majority of these
detections (92) are newly-identified class I methanol maser sources,
thus demonstrating that there is a high detection rate (55\%) of
class I methanol masers toward EGOs. This search, combined with
previous observations increased the number of known class I methanol
masers to $\sim$300. Chambers et al. (2011) obtained an apparently
contradictory result for a similar search, achieving a relatively
low detection rate (8/31=26\%) of class I methanol maser at 44 GHz
towards 4.5 $\mu$m emission sources.  The low detection rate in this
survey may be because Chambers et al. have targeted sources with
relatively less extended green emission than the EGOs identified by
Cyganowski et al. (2008).

The generally held view of class I methanol masers is that they
trace regions of mildly shocked gas, where the methanol abundance is
significantly enhanced and the gas is heated and compressed
providing more frequent collisions. Voronkov et al. (2010a)
suggested that the shocks which produce class I methanol masers may
be driven into molecular clouds not only by outflows (it is worth
noting that a high-velocity feature from a class I methanol maser
associated with outflow parallel to the line of sight has been
detected in the EGO source G309.38-0.13 by Voronkov et al. 2010b),
but also from expanding H{\sc ii} regions. Based on the results of
their analysis of GLIMPSE properties and the findings of Voronkov et
al., Chen et al. (2011) suggest that class I methanol masers may
arise at two distinct two-evolutionary phases during the high-mass
star formation process: they may appear as one of the first
signatures of massive star formation associated with young outflows,
and also that they can be re-activated at a later evolutionary stage
associated with OH masers and H{\sc ii} regions.

Further searches for class I methanol masers are very important for
our understanding of the range of environments and circumstances in
which they arise. Ellingsen (2006) developed criteria for targeting
class II methanol maser searches using GLIMPSE point source colors.
He suggested that targeted searches toward GLIMPSE point sources
with [3.6]-[4.5] $>$ 1.3 and an 8.0 $\mu$m magnitude less than 10
will detect more than 80\% of all class II methanol maser sources
with an efficiency of greater than 10\% (although the actual
efficiency obtained from the only follow-up search reported to date
is much lower (Ellingsen 2007)). In comparison, the mid-IR color
analysis of GLIMPSE point sources toward EGOs undertaken by Chen et
al. (2011) shows that the color-color region occupied by the GLIMPSE
point sources towards EGOs which are, and are not, associated with
class I methanol masers are very similar, and mostly located within
color ranges -0.6$<$[5.8]-[8.0]$<$1.4 and 0.5$<$[3.6]-[4.5]$<$4.0.
This suggests that the GLIMPSE point source colors may not be a very
sensitive diagnostic for constructing a sample to search for class I
methanol masers.  Despite the significant overlap in the color space
occupied by EGOs with and without an associated class I methanol,
Chen et al. (2011) do find the detection rate of class I methanol
masers is higher in those sources with redder GLIMPSE point source
colors. Therefore the reddest GLIMPSE point sources may provide a
reasonable sample for searching for class I methanol masers with a
relatively high detection efficiency. One point to note is that the
implication of a relatively high detection efficiency for class I
methanol masers for the redder GLIMPSE point sources is based on the
EGO sample. The GLIMPSE point sources associated with EGOs are
believed to be MYSOs with ongoing outflows, and the EGOs themselves
have a high detection rate of class I methanol masers. Therefore a
relatively high detection efficiency of class I methanol masers is
not unexpected in these redder GLIMPSE point sources associated with
EGOs. Further searches for class I methanol masers toward non-EGO
associated GLIMPSE point sources is necessary to more reliably
characterise the mid-IR characteristics of class I methanol maser
sources. The mid-IR colors of some other astrophysical objects,
(e.g. AGB stars) also are located within a similar color-color
region as that found for class I methanol masers (Robitaille et al.
2008). Thus finding additional measures by which GLIMPSE point
sources associated with active star formation can be distinguished
from other objects with similar mid-IR colors is an important step
required for such searches.

Recently the Bolocam Galactic Plane Survey (BGPS) has detected 1.1
mm thermal dust emission from thousands of regions of dense gas,
many of which are closely associated with star formation. The
typical H$_{2}$ column density of BGPS sources is $\sim10^{22}$ cm
$^{-2}$, the typical mass a few hundred M$_{\odot}$, and the typical
size a parsec (Aguirre et al. 2011; Dunham et al. 2011a, b). So the
BGPS is identifying high column density regions and is a sensitive
tracer of massive clumps, in contrast to signposts such as class II
methanol maser emission, which are only present once a YSO has
formed. Dunham et al. (2011a) found that approximately half the BGPS
sources contain at least one GLIMPSE source (within the area where
both BGPS and GLIMPSE surveys overlap). Chen et al. (2011) found
that the detection rate of class I methanol masers is significantly
higher towards those EGOs with an associated BGPS source
(35/54=65\%) than for those without (1/9=11\%), or in the complete
EGO sample (55\%). Dunham et al. (2011a) also found that EGOs are
frequently associated with BGPS sources.  Of the 84 EGOs within the
BGPS survey area, 79 are associated with BGPS sources. All of the
above factors suggest that the BGPS may be a useful supplement to
the GLIMPSE point source catalog in constructing a reliable and
efficient targeted sample for class I methanol masers.

In this paper, we report the results of a 95 GHz class I methanol maser survey
towards a sample of GLIMPSE point sources with associated 1.1 mm
BGPS sources which has been undertaken with the Purple Mountain
Observatory (PMO) 13.7 m radio telescope. In Section 2 we describe
the sample and observations, in Section 3 we present the results of
the survey, analysis and discussion is given in Section 4, followed
by a summary of the important results in Section 5.

\section{Source selection and Observations}
\subsection{Source selection}

We used the released catalogs from the GLIMPSE survey (version 2.0)
and the BGPS (version 1.0.1) to construct a target sample for our
class I methanol maser search. The properties of the two surveys are
summarized below.  The BGPS \footnote{See
http://irsa.opac.caltech.edu/data/BOLOCAM$_{-}$GPS} is a 1.1mm
continuum survey of 170 square degrees of the Galactic Plane in the
northern hemisphere with the Bolocam instrument (Glenn et al. 2003;
Haig et al. 2004), employed on the Caltech Submillimeter Observatory
(CSO). Two distinct portions were included in the survey: a blind
survey of the inner Galaxy region spanning $-10\arcdeg < l <
90\arcdeg$ where $|b|\leq0.5\arcdeg$ everywhere, except for
$1.0\arcdeg$ cross-cuts at $l=3\arcdeg$, $15\arcdeg$, $30\arcdeg$,
and $31\arcdeg$ where $|b|\leq1.5\arcdeg$, and a targeted survey
towards known star formation regions in several outer Galaxy
regions, including Cygnus-X ($70\arcdeg < l < 90.5\arcdeg$,
$|b|\leq1.5\arcdeg$), the Perseus Arm ($l \sim 111\arcdeg$,
$b=0\arcdeg$), the W3/4/5 region ($l \sim 135\arcdeg$,
$b\sim0.5\arcdeg$), IC1396 ($l \sim 99\arcdeg$, $b\sim3.5\arcdeg$)
and the Gemini OB1 molecular cloud ($l \sim 190\arcdeg$,
$b\sim0.5\arcdeg$). The survey detected approximately 8400 sources
with an rms noise level in the maps ranging from 30 to 60 mJy
beam$^{-1}$. The details of the survey methods and data reduction
are described in Aguirre et al. (2011), and the source extraction
algorithm and catalog (v1.0 BGPS data) are described in Rosolowsky
et al. (2010). The effective FWHM beam size of the BGPS is
33$\arcsec$, corresponding to a solid angle of $2.9\times10^{-8}$ steradians,
which is equivalent to a tophat function with a 40$''$ diameter
($\Omega=2.95\times10^{-8}$). Thus the BGPS catalog presents
aperture flux densities within a 40$\arcsec$ diameter aperture
($S_{40\arcsec}$), corresponding to the flux density within one
beam. The BGPS catalog also provides an integrated flux density
($S_{int}$), which is the sum of all pixels within a radius ($R$
also given in the catalog) of the BGPS source. Dunham et al. (2010)
suggested that a correction factor of 1.5 must be applied to the
Rosolowsky et al. BGPS catalog flux densities. This factor is based
on a comparison of BGPS data with previous 1.2 mm data acquired with
the MAMBO and SIMBA instruments (Aguirre et al. 2011). In this
paper, we have also applied this correction factor to the flux
densities in the Rosolowsky et al. BGPS catalog. The BGPS catalog
includes the coordinates of both a geometric centroid and of the
peak of the 1.1 mm emission. We have used the peak positions for the
dust continuum emission (rather than centroid positions) in our
analysis.

The Galactic Legacy Infrared Mid-Plane Survey Extraordinaire
(GLIMPSE)
\footnote{http://irsa.ipac.caltech.edu/data/SPITZER/GLIMPSE/} is a
legacy science program of the \emph{Spitzer Space Telescope} in a
number of mid-infrared wavelength bands at 3.6, 4.5, 5.8, and 8.0
$\mu$m using the Infrared Array Camera (IRAC; Benjamin et al. 2003;
Churchwell et al. 2009). The survey resolution is better than 2$\arcsec$ in
all wavelength bands. The survey catalogs for GLIMPSE I and II have
been released. The GLIMPSE I survey covers $10\leq|l|\leq65\arcdeg$
with $|b|\leq1\arcdeg$, and the GLIMPSE II survey covers the region
of $|l|\leq10\arcdeg$ with $|b|\leq1\arcdeg$ for $|l|>5\arcdeg$,
$|b|\leq1.5\arcdeg$ for $2\arcdeg<|l|\leq5\arcdeg$, and
$|b|\leq2\arcdeg$ for $|l|\leq2\arcdeg$. The data products include
both highly reliable point source catalogs, and less reliable
but more complete point source archives. In our analysis, we have
used only the highly reliable point source catalogs from the GLIMPSE
I and II surveys. The total area of the GLIMPSE I and II surveys is
274 square degrees. The overlap region between the BGPS and the two
GLIMPSE surveys is $-10\arcdeg < l < 65\arcdeg$, $|b|<0.5\arcdeg$
and $|b|<1.0\arcdeg$ at $l=3\arcdeg$, $15\arcdeg$, $30\arcdeg$, and
$31\arcdeg$. We have used data from the overlap region to compile a
sample of target sources to search for class I methanol masers.

The target sample was constructed by applying the following
criteria: (1) A GLIMPSE point source with [3.6]-[4.5]$>$1.3,
[3.6]-[5.8]$>$2.5, [3.6]-[8.0]$>$2.5 and an 8.0 $\mu$m magnitude
less than 10; (2) the point sources meeting this mid-IR criterion
must have a 1.1 mm BGPS counterpart within $15\arcsec$ (half the
beam size of the BGPS survey); (3) the source must be at a
declination greater than $-25\arcdeg$ (so as to be accessible to the
PMO 13.7-m telescope); (4) the separation of each target source from
all other target sources must be greater than half the beam size of
the PMO 13.7-m telescope at 95 GHz (30$''$) (where this is not the
case the source with stronger 4.5 $\mu$m emission is retained in the
sample). The mid-IR criteria for selecting the initial sample of
GLIMPSE point sources are based on the observed colors of known
class I and class II methanol masers (see Figures 15, 16 and 18 of
Ellingsen (2006)). Although Chen et al. (2011) found that some class
I methanol masers are associated with GLIMPSE point sources with
less-red colors ([3.6]-[4.5]$\sim$0.5), the detection rates are
highest for redder colors and so these criteria should be more
efficient.  When cross-matching the GLIMPSE point sources and the
BGPS sources we only considered the separation between the GLIMPSE
point source position and the BGPS peak position.  We did not
consider the measured size of the BGPS source, even though this
method may miss some true associations between GLIMPSE and BGPS
sources.

Within the BGPS-GLIMPSE overlap regions a total of $\sim$420 GLIMPSE
point sources satisfied the four criteria outlined above. Of these a
total of 214 (approximately half) were randomly picked as the
initial target sample for our 95 GHz class I methanol maser survey
with the PMO 13.7-m telescope. Table 1 lists the target sample
source parameters including the mid-IR magnitudes of the GLIMPSE
point sources and the main parameters of the BGPS sources (including
the BGPS ID number) extracted from the relevant catalogs used in
this study. The separation between the GLIMPSE point source and the
BGPS source range from 0.3$''$ to 14.7$\arcsec$ with a mean of
7.3$\arcsec$. A histogram of the separations is shown in Figure 1.

\subsection{Observations}

The observations of the $8_{0}$ -- $7_{1}$ A$^{+}$ (95.1964630 GHz)
class I methanol maser transition were made using the PMO 13.7 m
telescope in Delingha, China during 2011 March -- April. We used the
position of the 1.1 mm BGPS source peak emission rather than GLIMPSE
point source as the target position for the observations. The
positions of the target sources in Equatorial Coordinates (J2000)
are given in Table 2. A new cryogenically cooled 9-beam SIS receiver
($3\times3$ with a separation of 174$\arcsec$ between the centers of
adjacent beams) was used for the observations. This receiver
operates in the 80--115 GHz band and the central beam of the 9 beam
receiver was pointed at the target position. The system temperature
for the observations was in the range 105--140 K, depending on the
weather conditions and the atmospheric absorption $\tau$ was
typically 0.15 -- 0.2. A Fast Fourier Transform Spectrometer (FFTS)
with 16384 spectral channels across a bandwidth of 1 GHz
(corresponding to a velocity range of $\sim$ 3000~ km s$^{-1}$) was
available for each beam during the observations. This gives an
effective velocity resolution of 0.19 km s$^{-1}$ for the 95 GHz
class I methanol masers. However we only searched for maser emission
over the velocity range from -200 to 200 km s$^{-1}$ to cover the
range of observed molecular gas in the Milky Way. Each source was
observed in a position-switching mode with off positions offset
10$\arcmin$ in right ascension. The pointing rms was better than
5$\arcsec$. The standard chopper wheel calibration technique was
applied to measure an antenna temperature, T$_{A}^{*}$ corrected for
atmospheric absorption. The FWHM beam size of the telescope is
approximately 53$\arcsec$ at this frequency with a main beam
efficiency $\eta_{mb}$ of 46\%. The antenna efficiency is 42\%, thus
resulting in a factor of 45 Jy K$^{-1}$ for conversion of antenna
temperature into flux density. The initial observations had an
on-source integration time of 10 mins for each of the 214 targeted
sources yielding a T$_{A}^{*}$ 1$\sigma$ noise level of about 20 mK
(corresponding to about 1.0 Jy) for each beam after Hanning
smoothing of the spectra. Then, depending on the intensity of any
detected emission we observed for an additional 10-20 minutes
(on-source) to improve the signal-to-noise (SNR) of the final
spectra. This yielded a typical rms noise level of 15 -- 20 mK in
the T$_{A}^{*}$ scale (corresponding to 0.7 -- 1.0 Jy) after Hanning
smoothing. The corresponding rms noise ($\sigma_{rms}$) for each
target source is summarized in Table 2.

The spectral data were reduced and analyzed with the GILDAS/CLASS
package. Although data from all 9-beams were recorded, the locations
of the 8 offset beams rotate with changing azimuth/elevation during
the observation, thus only data from the central beam which was
placed on the target position is valid. We only focus on the data
from the central beam in this work. As part of the processing a
low-order polynomial baseline fitting and subtraction, and Hanning
smoothing were performed for the averaged spectra. Usually the 95
GHz methanol spectra do not have a particularly Gaussian profile,
possibly because the spectra consist of multiple maser features
within a similar velocity range. However, to characterize the
spectral characteristics of the emission we have performed Gaussian
fitting of each feature for each detected source.

\section{Results}

\subsection{Class I methanol maser detection}

95 GHz class I methanol emission above 3 $\sigma_{rms}$ was detected
toward 63 of the 214 targeted sources, corresponding to a detection
rate of 29\% for this survey. The spectra of the 63 detected class I
methanol sources are shown in Figure 2. The detected objects are
listed, along with the parameters of Gaussian fits to their 95 GHz
spectral features in Table 3. The flux densities of the detected
emission derived from the Gaussian fits range from $\sim$ 0.6 to
43.4 Jy (corresponding to main beam temperatures T$_{BM}$ $\sim$
0.03 to 2.1 K). The flux densities obtained from integrating the
emission over all spectral features for each source are also given
in Table 3 and range from 3 to 136 Jy km s$^{-1}$, with a mean of 24
Jy km s$^{-1}$. The measured FWHM of individual spectral features
derived from Gaussian fitting are in the range 0.18 -- 11.5 km
s$^{-1}$ with a mean of 2.1 km s$^{-1}$. The spectra of the class I
methanol emission in most sources usually include one or more narrow
spectral features (typical line width $<$1 km s$^{-1}$ seen in Table
2) which are clearly maser emission (see Figure~2), but the same
spectra often also contain broader emission features (typical line
width $>$1 km s$^{-1}$ seen in Table 2). The pattern of class I
methanol transitions containing both strong narrow spectral features
and weaker broader emission has been seen in all previous
single-dish surveys (e.g. Ellingsen 2005 and Chen et al. 2011), and
their nature was discussed in detail by Chen et al. (2011). There
are 13 sources which show a single broad Gaussian profile with a
width of $>$ $\sim$2 km s$^{-1}$ (sources N20, N29, N32, N41, N78,
N94, N99, N102, N117, N148, N154, N164 and N194; see Figure 2). At
present our single-dish observations can not distinguish from their
characteristics whether these broader emission sources are maser or
thermal. For the purposes of our subsequent analysis we have assumed
that some of the detected 95 GHz emission in these single broad line
sources arises also from masers, recognising that future
interferometric observations are required to determine whether or
not this is correct. One point to note is that even if these single
broad line sources are found to be purely thermal, the number (only
13) of these sources is too small to affect most of the statistical
conclusions drawn in Section 4.

\subsection{Comparison with previous detections}

Among the 63 detected 95 GHz methanol sources, 20 have previously
been detected as class I methanol masers in one or more transitions.
The previous class I maser observations of these 20 sources are
summarized in Table 4, including information as to which transitions
have been detected. Table 4 shows that 12 of these sources (all of
them are EGOs) have previously been detected in the 95 GHz
transition, 11 of them by Chen et al. (2011) (Mopra EGO survey) and
the other one from the survey of Val'tts et al. (2000). Twelve
sources were also detected in the 44 GHz transition, including 6
EGOs detected by Cyganowski et al. (2009) and Slysh et al. (1994) as
well as 6 sources from other surveys. Therefore 51 new 95 GHz class
I methanol maser sources have been found in this survey, of which 43
are newly-identified as class I methanol maser sources. One source
(source number N22 in our survey) was detected as a class I methanol
maser at 44 GHz, but undetected at 95 GHz in a survey with the
Nobeyama 45-m telescope by Fontani et al. (2010). While in our
observations we detected emission in the 95 GHz transition with a
peak flux density of $\sim$12 Jy. We have compared the targeted
positions for this source in the two surveys, and found that there
is an angular separation of $\sim18\arcsec$ between the targeted
positions used. If the 95 GHz methanol maser emission detected in
our PMO-13.7 m observations is located close to our targeted
position, the non-detection with the Nobeyama 45-m telescope may be
due to the relatively smaller beam size at 95 GHz ($18\arcsec$)
which may not have covered the maser emission region in this case.

We have compared the spectra of the 11 sources which were detected
in both the EGO-based Mopra survey (Chen et al. 2011), and in the
current PMO 13.7-m survey. The two spectra overlayed are shown in
Figure 3 and it can be clearly seen that the line profile and
velocity range of each source are similar in both surveys. The
observed emission intensities are consistent in 4 sources (N43, N73,
N76 and N83), but are different in other 7 sources. Usually the
emission detected in the current PMO survey is (1.5 -- 2 times)
stronger than that in the previous Mopra survey (except for one
source N97 with stronger emission detected in the Mopra survey). In
addition to flux density calibration uncertainties between the two
telescopes, the following factors may cause the observed difference
in the detected emission intensity between the two surveys: 1)
different target positions were adopted in the two surveys; 2) the
different beam sizes of the telescopes used in the two surveys cover
different regions; 3) intrinsic intensity variability in the class I
methanol maser emission between the two epochs. We have compared the
targeted positions used in the two surveys, and found that the
angular separation typically ranges from 1$\arcsec$ to 10 $\arcsec$
in both the sources with and without a significant difference in the
observed intensity, thus it does not seem that case 1 is the major
factor in explaining the differences. Case 2 is plausible if the
maser emission is extended to spatial scales comparable to, or
larger than the Mopra beam (36$\arcsec$ at 95 GHz), in which case
the PMO would detect additional maser emission outside the Mopra
beam. This is consistent with the observed results in the two
surveys for most sources, as stronger emission was detected by the
PMO, but one source (N97) shows the opposite trend with stronger
emission detected by the Mopra rather than the PMO. In this case one
of possibilities is that there is intrinsic intensity variability in
this source, although we can not characterise the nature of the
variations with only two epochs of data collected using different
telescopes. Moreover the exact coordinates of this source are
unknown, so it is possible that both in the Mopra and PMO
observations are at an offset position, in which case even a fairly
small difference in the telescope pointing of about 10$\arcsec$ can
lead to a higher intensity observed with a narrower beam (Mopra)
than with a broader beam (PMO), provided that Mopra was pointing
more directly towards the source. Variations in the intensity of 6.7
GHz class II methanol masers have been detected  with timescales on
the order of days to years (e.g. Goedhart et al. 2004; Ellingsen
2007; Goedhart et al. 2009; Szymczak et al. 2011) and some sources
have been found to exhibit periodic variability (e.g. Goedhart et
al. 2009; Szymczak et al. 2011). Intensity variation in class I
methanol masers has also been reported in a few sources (e.g. Kurtz
et al. 2004; Pratap et al. 2007), but to date there are no
systematic observations of class I methanol maser variability. It
will be necessary to perform multi-epoch observations with accurate
calibration to determine the characteristics of the intensity
variations in class I methanol masers.

\subsection{Distance and luminosity of class I methanol masers}

The distance and the distance-dependent integrated maser luminosity
for each of the 63 detected methanol maser sources are given in
Table 5. We used the Galactic rotation model of Reid et al. (2009),
with the Galactic constants set to, R$_{\odot}$= 8.4 kpc and
$\Theta_{\odot}$= 254 km s$^{-1}$ to estimate the distances. Since
class I methanol maser emission is generally observed to lie close
to the V$_{LSR}$ as measured from the thermal gas (e.g. Cyganowski
et al. 2009), the velocity of the brightest feature in the 95 GHz
maser spectrum was used in the distance calculation. All Galactic
rotation models suffer from ambiguity (known as kinematic
distance ambiguity) for sources which lie within the solar circle.
With the exception of the velocity associated with the tangent
point, there are two distances (referred to as a near and far
distance), either side of the tangent point, which will produce the
same line-of-sight velocity. All of the sources with 95 GHz methanol
masers detected in our survey fall within the solar circle. Where
present, an association between the detected class I methanol maser
source and an infrared dark cloud (IRDC) may allow us to resolve the
distance ambiguity. IRDCs are believed to represent sites where the
earliest stages of massive star formation are present (e.g. Egan et
al. 1998; Carey et al. 1998, 2000; Simon et al. 2006a, 2006b). They
are observed in absorption against the diffuse infrared background
especially at 8.0 $\mu$m, and hence the identification of IRDCs is
greatly biased toward nearby sources (and hence the near kinematic
distance), where they will show greater contrast against the diffuse
IR background (see Jackson et al. 2008). We have cross-matched the
63 detected 95 GHz methanol masers with the catalog of IRDCs seen in
the \emph{Spitzer} GLIMPSE images (Peretto \& Fuller 2009), and we
have undertaken visual inspection of the GLIMPSE 8 $\mu$m images for
those sources with $|l|<10\arcdeg$ (which are not included in
Peretto \& Fuller catalog). The information as to whether the class
I methanol maser detections are associated with IRDCs or not is
summarized in Column (8) of Table 5. We found 33 of 63 maser sources
for which the associated BGPS sources are spatially coincident and
structurally similar to IRDCs. We have assumed that these 33 sources
are at the near kinematic distance. The remaining 30 class I maser
sources are associated with BGPS which are not coincident with
IRDCs, and for these we have adopted the far kinematic distance.

To examine how reasonable (or otherwise) the above distance
assumptions are, we have cross-checked our distance determinations
for a subsample of class I maser sources for which the distance
ambiguity has been resolved in other studies. Some of our detected
class I maser sources have a class II methanol maser association
(see Section 4.3 for the identification of the class II maser
associations), and some of these have had the distance ambiguity
more directly resolved using HI self-absorption (HISA) from the
Southern Galactic Plane Survey (SGPS) or the VLA Galactic Plane
Survey (VGPS) by Green \& McClure-Griffiths (2011). We found that 9
of the 10 sources with IRDC associations (which we assume to be at
the near distance) are assigned the near distance by Green \&
McClure-Griffiths (2011), and 5 out of the 6 sources without IRDCs
(which we assume to be at the far distances) are assigned to be at
the far distance by their work. We have marked these sources with a
``G'' in Table 5, and adopted the distances from their work in our
analysis for these sources. Moreover some of our detected class I
maser sources are included in the sample of BGPS sources studied
with molecular lines (e.g., NH$_{3}$, HCO$^{+}$ and N$_{2}$H$^{+}$)
by Dunham et al. (2011b) and Schlingman et al. (2011). There are 16
sources (marked by ``S'' in Table 5) which are contained in the BGPS
sample with distances determined in Table 5 of Schlingman et al.
(2011). Among them 14 sources have distance solutions determine from
Galactic rotation (the other two sources N194 and N210 have no
reliable distance estimations from the Galactic rotation; see
below), and our distance determinations with the IRDC method for
them (including 13 sources with IRDC associations at near distance,
and 1 source without IRDC associations at far distance) are
consistent with that determined in Schlingman et al. (2011). The 9
sources (marked by ``D'' in Table 5) are included in the BGPS sample
with distances determined in Table 6 of Dunham et al. (2011b).
Comparing their distances with those estimated from our analysis on
the basis of the presence or otherwise of an IRDC (7 sources with
and 2 sources without IRDCs, respectively) are also generally
consistent with those estimated by Dunham et al. (2011b). In
addition, one point to note is that the identification of an IRDC
depends on the presence of a bright 8 $\mu$m infrared background, so
a source at the near distance without a significant infrared
background might be not identified as IRDC. Therefore for those
sources without an identified IRDC, the distance may be less certain
and biased toward large distances. The reliability of the distance
determinations for our sources without IRDC associations could
potentially be improved through additional HISA investigations,
however, Dunham et al. (2011b) find that HISA is unlikely to be
present for BGPS sources without an associated IRDC. They find that
for 215 BGPS sources without IRDC identifications listed in Table 6
of Dunham et al. (2011b), only 26 present a definite HISA features.
Hence, we have not undertaken any additional HISA determinations
beyond those already available in the literature, as the available
cross-checks show that our assumption of the near and far kinematic
distances for sources with and without IRDCs respectively appear
reasonable. The accuracy of this discriminator for kinematic
distances can't be accurately assessed with such a small sample,
however, if our results are representative then it is at $\sim$90\%.
In some cases the Galactic rotation model is not able to provide a
reliable distance estimate and for these sources (sources N33, N194
and N210 in Table 1) we have adopted a distance of 4 kpc for source
N33 (which has an IRDC association), and that determined by
Schlingman et al. (2011) for the other two sources N194 and N210.

Based on the estimated distances, the integrated luminosity of 95
GHz methanol maser, L$_m$ can be calculated from
L$_{m}$=4$\pi$$\cdot$D$^2$$\cdot$S$_{int}^m$, where $D$ is the
estimated distance and S$_{int}^m$ is the integrated flux density of
the 95 GHz emission. This assumes that maser emission is isotropic,
which is known to be false, however, in the absence of any
information on the beaming angle of the maser emission, nor our
alignment with respect to it, this is the only feasible approach
that can be undertaken.

\section{Analysis and Discussion}

\subsection{Mid-IR characteristics of GLIMPSE point sources}

Analysis of the mid-IR colors of GLIMPSE point sources associated
with EGOs with and without class I methanol maser detections has
been performed by Chen et al. (2011). No significant difference in
the mid-IR colors was found between the GLIMPSE point sources with
and without class I methanol masers in the EGO sample (see Figure 5
of their work).  We have performed the same analysis for our
observing sample to further investigate the mid-IR characteristics
of the GLIMPSE point sources which are, and are not, associated with
class I methanol masers. Although the detected class I maser sources
in our PMO survey include 12 EGOs which were considered in the color
analysis by Chen et al. (2011), the remaining 51 newly-discovered 95
GHz class I methanol maser sources (which includes 3 EGO associated
sources previously only detected in the 44 GHz transition) allows us
to explore in a more unbiased manner, the mid-IR characteristics of
GLIMPSE point sources associated class I methanol masers.

A number of color-color diagrams were constructed to compare the
mid-IR colors of the GLIMPSE point sources with and without an
associated class I methanol maser detection in our survey. In Figure
4 we plot three color-color diagrams ([3.6]-[4.5] vs. [5.8]-[8.0];
[3.6]-[5.8] vs. [3.6]-[8.0] and [3.6]-[4.5] vs. [4.5]-[8.0]) using
different symbols for the sources which are, and which are not
associated with class I methanol masers. There are 63 members of the
group associated with class I methanol masers and 151 members of the
group which are not associated with class I methanol masers. This
figure shows that there are no clear differences in the mid-IR
colors between those sources in our sample which are associated with
a class I maser, and those which are not, consistent with the
findings from the EGO-based sample of Chen et al. (2011). There
are 15 sources in total associated with known EGOs in our observing sample
(see Table 4). The color regions occupied by the sources at
evolutionary Stages I, II and III, (derived from the 2D radiative
transfer model of Robitaille et al. (2006)), are marked on the
[3.6]-[4.5] vs. [5.8]-[8.0] color diagram of Figure 4 (left panel).
We found that most (187/214) sources in our observed sample fall in
the region occupied by the youngest protostars (Stage I), with the
remaining 27 sources found in the upper-left of the color-color
diagram, outside the Stage I evolutionary region. Chen et al. (2011)
have discussed these redder GLIMPSE sources which lie outside the
Stage I color region in detail. They may be deeply reddened sources
(with reddening vector A$_v\sim$80; a typical reddening vector of
A$_v$$=$20 derived from the Indebetouw et al. (2005) extinction law
is shown in Figure 4 to demonstrate the reddening effect), MYSOs
with an extremely high mass envelope, or caused by emission
mechanisms such as H$_{2}$ or PAH line emission which were not
included in the Robitaille et al. (2006) models. One of the most
likely explanations is that they have excess 4.5 $\mu$m emission
from shocked H$_{2}$ in particularly strong/active outflows, which
in turn readily produces class I maser emission. This is supported
by the high detection rate of class I methanol masers towards these
redder sources seen in both the current observations (17/27=63\% in
this survey), and the EGO survey of Chen et al. (2011) (a detection
rate of 75\%). We discuss possible dependence of the detection rate
of class I methanol masers with the colors or magnitudes of GLIMPSE
point sources in greater detail in Section 4.4.  For the redder
GLIMPSE point sources (outside the Stage I region), 8 sources with
an associated class I methanol maser are also associated with EGOs
(marked by red triangles in Figure 4), which means the other 9
sources with an associated class I masers are not associated with an
EGO, although 3 of them are associated with known MYSOs (sources
N22, N90 and N101; see Table 4).

We have undertaken a detailed analysis of possible correlations
between the class I methanol maser emission and the associated
GLIMPSE point sources. Figure 5 (left panel) shows a log-log plot of
the integrated luminosity of the class I methanol masers versus the
luminosity of the GLIMPSE point sources in the 4.5 $\mu$m band. The
distance to the source listed in Table 5 was used to calculate the
luminosity for both the class I maser and the GLIMPSE point source
(see the discussion of distance assignment in Section 3.3). A linear
regression analysis for this distribution was undertaken, and the
line of best fit obtained is plotted in the figure. Our analysis
suggests that there is a statistically significant positive slope in
the distribution, but with a weak correlation (the best fit shows a
slope of 0.41 with a statistically significant p-value of 10$^{-4}$
which allows us to reject zero slope in the data, and a low
correlation coefficient of 0.47). Such a correlation seems
reasonable if the 4.5 $\mu$m emission is believed to be enhanced by
shocks, which are also thought to be responsible for the class I
methanol maser emission. On the other hand, this correlation may be
simply a consequence of the correlation between the class I methanol
maser and central source luminosity, which has been obtained by,
e.g., Bae et al. (2011) for the 44 GHz masers. However, our
determination of the distances using the presence or absence of an
IRDC to resolve the distance ambiguity will introduce unpredictable
uncertainties as discussed in Section 3. To eliminate the possible
effects of distance dependencies in our investigations we compared
mid-IR color [3.6]-[4.5] with [3.6]-log(S$_{m}$), where S$_{m}$ is
the integrated flux density of the class I methanol maser (a plot of
this is shown in the right-hand panel of Figure 5). This plot shows
no significant correlation between the ``colors'', with the linear
regression analysis giving a slope of 0.6, a non-significant p-value
of 0.10 and a small correlation coefficient (0.22). One possible
reason for weak or non-significant correlation between them is that
the GLIMPSE point sources which have been identified as being
associated with the class I methanol masers may not be the true
driving sources. Within the large field-of-view covered by the PMO
beam size (52$\arcsec$), there will always be a number of GLIMPSE
point sources, and from the present observations with this
resolution we can not determine which one is the driving source of
the class I methanol maser. Our assumption that the GLIMPSE point
source which satisfies the mid-IR color criteria for the class I
maser search in our survey is the driving source is almost certainly
wrong in some cases, indeed some driving sources of class I methanol
maser are likely not present in the GLIMPSE point source catalog due
to saturation, the presence of bright diffuse emission, or
intrinsically extended morphology in the IRAC bands (e.g. from
extended PAH emission or extended H$_{2}$ emission in shocked gas
(see Robitaille et al. 2008, Povich et al. 2009, and Povich \&
Whitney 2010)). On the other hand, if the GLIMPSE point sources do
correspond to the true driving sources of the class I methanol
masers, the lack of significant correlations between the maser and
GLIMPSE mid-IR colors suggests that the excitation of the class I
methanol masers are not directly related to the mechanism
responsible for the mid-IR emission. This view is supported by the
fact that class I methanol maser spots are often distribute over
large angular and spatial scales (usually of the order of
10$\arcsec$), and are excited in shocked regions (e.g. Cyganowski et
al. 2009), whereas the GLIMPSE point sources emission reflects the
thermal dust or molecular environments within a smaller region
around the protostar. Moreover, the 4.5 $\mu$m emission may still be
dominanted by the thermal dust emission from the driving protostar,
rather than the molecular gas (such as H$_{2}$ or CO) excited by
shocks, thus masking any relationship between the class I methanol
maser properties and the 4.5 $\mu$m intensity.

\subsection{Relationships between class I methanol masers and BGPS sources}

A close correlation between GLIMPSE point sources with an associated
class I methanol masers and the presence of a 1.1 mm BGPS sources
was first noted in the EGO-based survey of Chen et al. (2011), the
analysis of which motivated the investigations undertaken here. Chen
et al. showed that the luminosity of the class I methanol masers in
the EGO sample strongly depends on the properties (including both
the mass and volume density) of the associated 1.1 mm dust clump:
the more massive and denser the clump, the stronger the class I
methanol emission. Here we perform a similar analysis to Chen et al.
(2011) on a sample of class I methanol masers which combines GLIMPSE
point sources and BGPS sources to investigate the relationship
between the dust clumps and the maser emission in a wider sample of
sources.

Based on the assumption that the 1.1 mm emission from the BGPS
source arises from optically thin dust, we can calculate the
associated gas mass using the equation:
\begin{equation}\label{1}
M_{gas}=\frac{S_{int}D^{2}}{\kappa_{d}B_{\nu}(T_{dust})R_{d}},
\end{equation}
where $S_{int}$ is the 1.1 mm integrated flux density of the BGPS
source, $D$ is the distance to the source, $\kappa_{d}$ is the mass
absorption coefficient per unit mass of dust, $B_\nu(T_{dust})$ is
the Planck function at temperature $T_{dust}$, and $R_d$ is the
dust-to-gas mass ratio. Here we have used $\kappa_{d}$$=$1.14
cm$^{2}$ g$^{-1}$ for 1.1 mm (Ossenkopf \& Henning 1994) and a
dust-to-gas ratio ($R_{d}$) of 1:100 in our calculations and
$B_\nu(T_{dust})$ was calculated for an assumed dust temperature of
20 K. The average H$_{2}$ column density ($N_{H_{2}}$) and volume
density ($n_{H_{2}}$) of each dust clump were then derived from its
mass and radius ($R$), assuming a spherical geometry and a mean mass
per particle of $\mu=2.37$ m$_{H}$. The parameters of the 1.1 mm
continuum integrated flux density, $S_{int}$ and 1.1 mm source
radius, $R$ were obtained from the BGPS catalog (Rosolowsky et al.
2010) for the 214 sources in our sample and the values are listed in
Table 1. We applied a correction factor of 1.5 to the Rosolowsky et
al. BGPS catalog flux densities (which are also listed in Column
(11) of Table 1 of our work) to derive the gas masses for the 63
BGPS sources with an associated class I methanol maser detection.
For the two sources (sources N39 and N143) which are unresolved with
the BGPS beam, we were not able to determine their gas column and
volume densities due to the absence of the size of the BGPS source.
The derived masses and gas densities for the 1.1 mm dust clumps with
an associated class I methanol maser are given in Table 5. As stated
in Section 3.3, there is a small number of detected class I maser
sources which are also included in the sample of BGPS sources
investigated by Dunham et al. (2011b) or Schlingman et al. (2011).
Comparing the physical parameters (gas mass and volume/column
density) derived for those BGPS sources which are in common with the
two previous studies, we find that they are consistent with each
other (usually similar but not identical).

A log-log plot of the luminosity of the class I methanol maser
versus the derived gas mass (left panel) and H$_{2}$ volume density
(right panel) of the associated 1.1 mm BGPS source is shown in
Figure 6. From this figure it can be seen that there is significant
positive correlation between the class I maser luminosity and the
gas mass of the BGPS source, while a very weak negative correlation
exists between the class I maser luminosity and the H$_{2}$ volume
density. Linear regression analysis for both distributions (the
corresponding best fit lines are overlaid in each panel of Figure 6)
find a statistically significant (p-value of 8.1E-13) linear
dependence with a slope of 0.81 existing between the maser
luminosity and the gas mass (the slope has a standard error of 0.07
and a correlation coefficient of 0.84).  In contrast, there is no
statistically significant correlation (p-value of 0.10) between the
class I methanol maser luminosity and the gas density (the fit has a
slope of -0.25 and a small correlation coefficient of 0.22). The
statistically-significant positive correlation between class I maser
luminosity and BGPS source mass obtained in this study is similar to
that measured in the EGO-based sample of Chen et al. (2011).  Chen
et al. (2011) also found a weak but statistically significant
positive correlation between the class I maser luminosity and the
gas volume density in the EGO sample, however, no statistically
significant or a very weak negative correlation is observed in our
larger and more diverse sample.

We also carried out an investigation of the dependence between the
BGPS beam-averaged gas column density and the class I methanol maser
integrated flux density (both of which are independent of the
assumed distance to the source). The H$_{2}$ column density per beam
can be estimated by
\begin{equation}\label{2}
N_{H_{2}}^{beam}=\frac{S_{40''}}{\Omega_{beam}\mu\kappa_{d}B_{\nu}(T_{dust})R_{d}},
\end{equation}
where S$_{40''}$ is the 1.1 mm flux density within an aperture with
a diameter of 40$\arcsec$, $\Omega_{beam}$ is the solid angle of the
beam, $\mu$ is the mean mass per particle, $\kappa_{d}$ is the mass
absorption coefficient per unit mass of dust, $B_\nu(T_{dust})$ is
the Planck function at temperature $T_{dust}=20$ K, and $R_d$ is the
dust-to-gas mass ratio, as described above. S$_{40''}$ was adopted
as the measure of the flux within a beam since a top-hat function
with a 20$\arcsec$ radius has the same solid angle as a Gaussian
beam with an FWHM of 33$\arcsec$ (see also Section 2.1). In addition
to a flux correction factor of 1.5 (see above), an aperture
correction of 1.46 should be applied to flux density S$_{40''}$
(which is given in Column (10) of Table 1 in our work) to account
for power outside the 40$\arcsec$ aperture due to the sidelobes of
the CSO beam (Aguirre et al. 2011) in the calculation of
beam-averaged column density. Since this property is independent of
the distance to the source, we can derive it for all BGPS sources in
our sample and we have listed it for each source in Column (12) of
Table 1. The results are shown as a log-log plot in Figure 7 which
demonstrates that there is a statistically significant positive
correlation between the beam-averaged gas column density of BGPS
sources and the integrated flux density of class I methanol masers
(S$_{int}^m$). We have performed a linear regression analysis for
this distribution and obtain a best fit linear equation of:
\begin{equation}\label{3}
log(S_{int}^m)=0.75[0.10]log(N_{H_{2}}^{beam})-15.94[2.28]
\end{equation}
with a correlation coefficient of 0.69 and p-value of 3.25E-10. This
relationship between the class I maser flux density and the
beam-averaged gas column is important for refining future class I
methanol maser surveys based on BGPS sources because it is
independent of distance and other intrinsic physical parameters of
the sources. For example, toward nearby low-mass star-forming
regions a threshold column density of 123 M$_{\odot}$ pc$^{-2}$
(corresponding to $6.5\times10^{21}$ cm$^{-2}$) has been observed
(Lada et al. 2010; Heiderman et al. 2010), and substituting this
into the above relationship we can estimate a lower limit of 2.6 Jy
km s$^{-1}$ for the integrated flux density of 95 GHz class I
methanol masers. The lowest class I maser integrated flux density
from our observations is only slightly higher ($\sim$ 3.0 Jy km
s$^{-1}$), which suggests that we are likely to have detected
significant part the 95 GHz class I maser sources in the observed
sample.

\subsection{Star formation activity associated with methanol masers}

The star formation activity of the BGPS sources was characterized by
Dunham et al. (2011a), through the properties of mid-IR sources
along a line of sight coincident with the BGPS sources. They divided
the BGPS sources into four groups representing increasing
probability of the associated mid-IR sources indicating star
formation activity. The sources with the highest probability of star
formation activity are classified as group 3 and include BGPS
sources matched with EGOs or Red MSX Survey (RMS; Hoare et al. 2004;
Urquhart et al. 2008) sources. The lowest probability group (group
0) includes BGPS sources which were not matched with any mid-IR
sources and are considered to be ``starless'' in their work.  Groups
1 and 2 represent BGPS sources matched with GLIMPSE red sources
cataloged by Robitaille et al. (2008), or a deeper list of GLIMPSE
red sources created by Dunham et al. (2011a). Overall they found
that the mid-IR emission associated with BGPS sources with a high
probability of star formation activity (group 3) are typically
extended with large skirts of emission, while the low probability
sources (group 1) are more compact, with weak emission. In this
section, we explore the star formation activity in the sources with
and without methanol maser associations using the parameters of the
BGPS sources.

Histograms of BGPS source parameters (beam-averaged H$_2$ column
density $N_{H_{2}}^{beam}$, integrated flux density $S_{int}$ and
radius R) for those sources with and without an associated class I
methanol maser are presented in Figure 8. Unfortunately we are not
able to compare any intrinsic physical parameters such as mass,
source size in pc etc between the two groups, due to the absence of
a distance estimate for the sources without an associated class I
methanol maser. For each distribution in Figure 8, the upper and
lower panels correspond to the BGPS sources with and without a class
I maser detection, respectively. It can be clearly seen that the
distributions differ significantly between BGPS sources with an
associated class I maser and those without for the beam-averaged
H$_2$ column density and the integrated flux density of BGPS sources
(see left-hand and middle panels). In contrast there is no
significant difference in the observed distribution of the radius of
the BGPS sources for the two samples (see right-hand panel). The
basic statistical parameters such as mean, median, standard
deviation, for each of these distributions are summarized in Table
6. The mean logarithm of the beam-averaged column density
$N_{H_{2}}^{beam}$ is 21.9 [cm$^{-2}$] for the sources with no class
I methanol maser detections, but 22.7 [cm$^{-2}$] for the sample of
sources with an associated class I methanol maser (a difference of
approximately 3 standard deviations). While the mean logarithm of
the BGPS integrated flux density is 0.0 [Jy] in sources without an
associated class I masers, but greater at 0.7 [Jy] in sources with
class I masers (a difference of approximately 2 standard
deviations). However, a {\em t}-test finds that the difference in
the mean of each of the distributions for these two properties is
statistically significant for the two groups. The distributions of
radii are not significantly different between the two groups, each
having a mean of around 50$\arcsec$ and a large range (mostly
distributed between 20 and 100$\arcsec$). The beam-averaged column
density for the BGPS sources without an associated class I maser
ranges between 21.4 [cm$^{-2}$] $\leq log(N_{H_{2}}^{beam})\leq$
22.7 [cm$^{-2}$], whereas for sources with an associated class I
masers the range is 21.9 [cm$^{-2}$] $\leq
log(N_{H_{2}}^{beam})\leq$ 23.8 [cm$^{-2}$]. Similarly the range of
the logarithm of integrated flux density is from -0.9 to 1.1 [Jy]
for BGPS sources without an associated class I masers, but from -0.6
to 1.5 [Jy] for those with a class I masers. The large overlapping
range in the integrated flux density distribution of the two groups
suggests that the beam-averaged column density is the most
efficacious parameter for selecting BGPS sources likely to be
associated with a class I methanol maser.

Comparing the distribution of the beam-averaged H$_{2}$ column
density for the four star formation activity groups described by
Dunham et al. (2011a; Figure 12), with that for the class I methanol
maser sample we can see that it is similar to that shown for group
3.  While the distribution for the sources without an associated
class I methanol maser is similar to that seen for group 0 and group
1 by Dunham et al. Comparing distributions of BGPS source flux for
our samples with Dunham et al. (2011a), those without class I masers
appear to agree well with their group 1. As described above, group 3
contains the sources with the highest probability of star formation
activity include BGPS sources matched with EGOs or RMS sources,
while group 0 represents those with the lowest probability,
including BGPS sources without any associated mid-IR source
(referred as ``starless''). Since all of our target BGPS sources
have an associated GLIMPSE point source we would expect that the
distributions we observe should differ from those seen for group 0
sources, which have no associated mid-IR source. The group 1
category sources include at least one IR object which may be an AGB
star catalogued by Robitaille et al.(2008) or a deeper GLIMPSE red
source from the list of Dunham et al. The BGPS sources in our sample
with an associated class I masers (63 in total) includes 15 EGOs
(which are classified into group 3 by Dunham et al.), however, the
relatively small number of EGOs can not dominant the BGPS parameter
distributions observed for this group. The remaining 48 sources must
also have a similar BGPS parameter distribution to that observed for
group 3. Since class I methanol maser emission is only known to be
found towards active star formation regions, the similar
distribution of the BGPS properties seen in the class I maser
sources and the group 3 sources supports the speculation of Dunham
et al. (2011a) that group 3 sources are those with the highest
probability of star formation activity. The BGPS sources in our
target sample without an associated class I methanol maser,
correspond to group 1 in the Dunham et al. classification (which
have a lower probability of star formation activity), and these may
be regions which are either too young, or have too low gas density,
or too weak outflows to excite class I maser emission.  Comparing
our observing sample with the GLIMPSE red source catalog complied by
Robitaille et al. (2008), we found that there are 95 sources are
common in the two data sets (including 22 sources with class I
masers and 73 sources without class I masers). Using the criteria of
Robitaille et al. to separate AGB stars and YSOs, 8 of 22 sources
for which we have detected an associated class I masers are
classified as extreme AGB stars with high mass-loss rates and
therefore significant circumstellar dust. However, since class I
methanol masers appear to only be associated with star formation,
this suggests that there may be a relatively high mis-classification
rate for the extreme AGB sources using the Robitaille et al.
criteria. We found that only 9 of 73 BGPS sources from our sample
which are not associated with class I masers are classified as AGB
stars. This also supports the hypothesis that the sources without
class I masers may be objects at early stages of star formation,
rather than AGB stars.

Analysis of the properties of 1.1 mm BGPS associated with EGOs by
Chen et al. (2011) showed that those which are associated with class
I methanol masers, but not class II methanol masers have a lower
mass/density of dust clump than those which are associated with both
class I and II methanol masers. We have cross-matched the 63 sources
with an associated class I methanol maser detected in our survey
with the catalog of 6.7 GHz class II methanol masers (usually better
than 1$''$) from the Parkes Methanol Multibeam (MMB) blind survey
published to date (Caswell et al. 2010; Green et al. 2010 ; Caswell
et al. 2011 ; Green et al. 2012), or from the ATCA observations of
Caswell (2009). The MMB masers positions have been measured to high
positional accuracy (better than 1$\arcsec$) and the observations
have a sensitivity of about 0.2 Jy ($3\sigma$ from the subsequent
ATCA observations).  The MMB survey published to date covers the
region  $186\arcdeg<l<20\arcdeg$ with $|b|<2\arcdeg$. Thus the
overlap region between the class I methanol maser sources detected
in our survey and the class II methanol masers in the MMB survey is
$0\arcdeg<l<20\arcdeg$ with $|b|<0.5\arcdeg$. The MMB survey data
from the overlap region allow us to identify the associations
between the two classes of methanol masers, and in particular to
identify those class I methanol maser sources without an associated
class II masers. Whether the class I methanol maser detected in this
survey is associated with a class II maser or not is summarized in
Column (9) of Table 5. Thirty three of the 63 class I methanol
masers in our sample lie in the MMB overlap region and Caswell
(2009) data set, and of these 20 have an associated class II maser
and 13 do not. Histograms of the beam-averaged H$_{2}$ column
density and flux density of BGPS for the sources associated with
only class I methanol masers compared to those associated with both
classes of methanol maser are shown in Figure 9. Although the sample
sizes for the two groups are relatively small, they still allow us
to investigate whether the BGPS properties discriminate between the
two groups. The statistical parameters for each distribution are
summarized in Table 6. From Figure 9 and the statistical parameters
we can see that there is a trend for the sources associated both
methanol maser classes to have higher BGPS flux densities and column
densities than the sources associated with only class I masers. The
mean column density and flux density of the associated BGPS sources
are marked with a dashed line in the corresponding histogram, and
are significantly larger for the sample of sources associated with
both classes of methanol maser. A {\em t}-test shows that the
difference in the mean of the two group distributions for the two
BGPS properties is statistically significant.

It is important to note that the sample size used in the current
analysis is small. A larger sample is required to more thoroughly
investigate the star formation activity and physical properties of
the regions with associated class I and II methanol masers. In
addition, our assumption of a dust temperature (T$_{dust}$) of 20 K
for all sources in our analysis will affect the physical parameters
such as mass and column/volume density derived from the BGPS data.
For example, a dust temperature of 7.2 K for the BGPS sources with
an associated class I methanol maser and a dust temperature of 20K
for those without would result in distributions of the beam-averaged
column density for the two samples having the same mean.  However,
the mean gas kinetic temperature derived from the NH$_{3}$
observations for group 3 sources (those similar to the class I maser
group) was 22.7 K (Dunham et al. 2011b), much higher than the 7.2 K
required to give the distributions the same mean.  Furthermore, since
the BGPS sources without an associated class I maser are similar to
group 1 of Dunham et al.,  for which the mean temperature from
NH$_{3}$ observations is 14.6 K (Dunham et al. 2011b), the
expectation is that more accurate temperature estimates for
individual BGPS sources would produce a greater difference in the
distributions of the physical properties derived from BGPS data,
rather than reducing it.

\subsection{Detection rates}

In this section, we compared the detection rates of class I methanol
masers with the cataloged parameters of the associated GLIMPSE point
sources and 1.1 mm BGPS sources with the aim of developing more
efficient criteria for future targeted class I methanol maser
searches.

Figure 10 presents a histogram showing the detection rate of class I
methanol masers as a function of the 4.5 $\mu$m magnitude (left
panel) and [3.6]-[4.5] color (right panel) of the associated GLIMPSE
point sources. It can be clearly seen that the detection rate for
class I masers increases (from 0.1 to 0.5) as the 4.5 $\mu$m
magnitude decreases (i.e. with increasing 4.5 $\mu$m flux density).
In contrast the detection rate for class I methanol masers shows no
significant variation for [3.6]-[4.5] color $<$ 3.2, being $\sim
0.2-0.3$, however for [3.6]-[4.5]$>$3.2 it is much higher
(0.8--1.0). Recalling the discussion in Section 4.1, these results
are consistent there being no significant differences between the
mid-IR colors of the sources with and without an associated class I
methanol maser, however, there is a higher detection rate for class
I methanol masers towards GLIMPSE point sources with the most
extreme red range for [3.6]-[4.5] color. Chen et al. (2011) have
suggested that these redder sources may correspond to higher mass,
high luminous YSOs. The correlation between the detection rate of
class I methanol masers and the 4.5 $\mu$m magnitude (or flux
density) of the associated GLIMPSE point source suggests that the
outflows or shocks are stronger or more active for those with more
intense 4.5 $\mu$m emission which thus are more likely to produce
maser emission. Apart from correlation between the emission
intensities of class I methanol masers and the GLIMPSE 4.5 $\mu$m
band (Figure 5, left), there is clearly an increased probability of
the presence of a methanol maser for sources with stronger 4.5
$\mu$m emission. This may be because although strong 4.5 $\mu$m
emission is a good indicator of the presence of shocks (and hence
the possibility of a class I maser), the intensity of the maser may
depend more strongly on other physical factors such as the gas mass
and column density of the parent clouds. We also note that while
sources with [4.5]$<$8.0 have a higher detection rate for class I
methanol masers (0.4 -- 0.5), they were typically classified as
extreme AGB stars by Robitaille et al. (2008).  At present all class
I methanol masers are thought to be associated with star formation,
which suggests that there is a high mis-classified rate for the
extreme AGB star population in Robitaille et al. (2008) and that
many of these sources correspond to luminous YSOs.

The detection rates of class I methanol masers as a function of the
BPGS cataloged parameters (beam-averaged H$_{2}$ column density and
integrated flux density) are shown in Figure 11. The BGPS radius
parameter is not included in the analysis because as discussed in
section 4.2, the radius is the least useful BGPS parameter in terms
of its ability to select BGPS sources with a higher likelihood of
having associated class I maser emission. From this figure, we can
clearly see that the detection rates for class I methanol masers
significantly increase with increasing values of the BGPS source
parameters. To allow a more detailed comparison the number and rate
of detection for class I masers in each bin for each BGPS parameter
in our observed sample and the full BGPS catalog are summarized in
Table 7. This shows that the detection rate for this sample is 100\%
for sources with the highest beam-averaged column density (larger
than 23.0 [cm$^{-2}$] in logarithm) and BGPS integrated flux density
(larger than 1.2 [Jy] in logarithm). We note that if there are
thermal sources among the objects detected at 95 GHz, they may be
preferentially associated with BGPS sources with the highest column
densities. As the number of these BGPS sources in the high column
density bins is small, even a few sources can potentially distort
the statistics. To test for this we excluded the 13 (potentially
thermal) sources with a single broad line profile (as identified in
Section 3.1), from the class I methanol maser detection sample and
from the total sample and redid our analysis. Our re-analysis
excluding potential thermal sources showed a similar trend to that
seen in Figure 11. In fact, among the 13 broad line profile sources,
only 3 are located in the high column density bins ($>10^{23}$
cm$^{-2}$) with 100\% probability of a 95 GHz maser detection. Thus
the possible thermal 95 GHz sources do not precisely correspond to
BGPS sources with the highest column densities.  The rate (3/9) of
the possible thermal sources to the class I methanol detection
sources in the high column density bins is relatively low, thus the
possible thermal sources do not significantly distort the
statistics. Moreover, as stated in Section 3.1, we can not exclude
the possibility that the emission from weak maser features
contributes to the broad line profiles. Our analysis using all 95
GHz detections does not exclude any possible maser sources for a
future survey toward a larger BGPS sample (see below). Even if we
assume that all the broad line profile sources are totally thermal,
the rate of real maser sources would still be very high (50/63=80\%)
in any sample derived on the basis of all 95 GHz detections.

For class I methanol maser surveys with a single dish with a beam
size of around an arcminute, it seems that the millimetre continuum
emission on similar scales (e.g. the BGPS sources) can provide a
better targeting criteria than the arcsecond-scale mid-IR emission
(e.g. GLIMPSE point sources). Our earlier discussion shows that the
class I methanol maser emission intensity is not closely related to
the mid-IR emission of GLIMPSE points sources, but does depend on
the mass and beam-averaged column density of the associated BGPS
sources, also suggesting that BGPS properties are likely to provide
a better basis for constructing samples for further class I methanol
maser searches. We also undertook binomial generalized linear
modeling (GLM) for the class I maser presence and absence using both
the GLIMPSE point source and BGPS properties, similar to that
undertaken for water masers by Breen et al. (2007) and Breen \&
Ellingsen (2011). This investigation showed that the BGPS source
properties are a much stronger predictor of the likelihood that a
particular source will host a maser, than are the mid-IR properties,
consistent with the investigations outlined above.  As the results
of the binomial GLM are less readily interpreted than the more
direct correlation investigations in sections 4.1, 4.2 and 4.3, and
don't reveal any significant new information we do not discuss them
further here.

To more efficiently search for class I methanol masers using the
BGPS sources, we can combine the two BGPS properties of beam-averaged
column density and integrated flux density to develop better
criterion for future searches. In Figure 12 (left panel) we plot a
log-log distribution comparing the BGPS flux density versus BGPS
beam-average column density from the current observations using
different symbols for the sources with and without class I methanol
maser detections (including also possible thermal sources). This
clearly shows that there is a significant discrimination between
sources with class I masers (marked by red circles) and those
without class I masers (marked by blue triangles). From inspection
of this plot we have defined a region wherein most (90\%) of class I
methanol maser detected in our current survey are placed,
constructed with red lines in the plot. The defined region can be
expressed as follows:

\begin{equation}\label{1}
  log(S_{int})\leq-38.0+1.72log(N_{H_{2}}^{beam}) ,
  and \ log(N_{H_{2}}^{beam})\geq 22.1 ,
\end{equation}
We can then extrapolate the identified class I methanol maser region
to the full BGPS sample to estimate the likely number of class I
methanol masers. The distribution of all BGPS sources with the class
I maser region overlaid is present in the right panel of Figure 13.
In total, approximate 1200 sources are located within the defined
class I maser region. If we extrapolate the results of this study we
would predict that we can detect 90\% of all the expected
($\sim1000$; see Table 7) class I methanol masers associated with
BGPS sources and that the detection efficiency would be about 75\%
towards the sources within the defined region. Since the above
estimates are based on all 95 GHz detections (including possible
thermal sources), the number of real maser source detections may be
at 80\% (at worst) of the above predictions (see discussion above).

However, a search for class I methanol maser towards an unbiased
sample of BGPS sources is required to clarify all possible sample
selection effects and to reliably estimate the true number of
methanol masers associated with the full BGPS catalog. Since the
BGPS only covers around half of the inner Galaxy, the total number
of class I methanol masers in the Galaxy would be expected to be at
least double the number associated with BGPS sources, suggesting
that class I methanol masers may be significantly more numerous in
the Galaxy than are class II methanol masers.

\section{Summary}

Using the PMO 13.7-m radio telescope, we have performed a search for
95 GHz class I methanol masers toward a sample selected from a
combination of the mid-IR \emph{spitzer} GLIMPSE and 1.1 mm CSO BGPS
surveys. A total of 214 sources were selected as the observing
sample, and these satisfy the GLIMPSE mid-IR criteria of
[3.6]-[4.5]$>$1.3, [3.6]-[5.8]$>$2.5, [3.6]-[8.0]$>$2.5 and 8.0
$\mu$m magnitude less than 10, and are also associated with a 1.1
mm BGPS source. 95 GHz class I methanol maser emission was detected
toward 63 sources, of these 51 are new 95 GHz class I methanol maser
sources, and 43 have no previously observed class I methanol maser
activity. Thus a detection rate of $\sim$29\% was observed for class
I methanol masers in the conjunct sample of GLIMPSE and BGPS surveys
from our single-dish survey. We also find that the sensitivity of
survey exceeds the theoretical detection limit derived from the
observed dependence between the integrated class I maser emission
and the BGPS beam-averaged column density.

Analysis of the mid-IR colors of GLIMPSE point sources in our
observing sample indicates that the color-color region occupied by
those sources with and without an associated class I methanol maser
are not significantly different. However, the detection rate of
class I methanol masers is higher towards those GLIMPSE point
sources with redder mid-IR colors. The mid-IR characteristics the
GLIMPSE sources associated with class I methanol masers in the
current sample is very similar to that derived in our earlier
EGO-selected sample. We find that the class I methanol maser
intensity is not closely related to either the mid-IR emission
intensity nor the color of the associated GLIMPSE point sources.
However, the maser emission is well correlated with the gas mass
derived for the BGPS sources. Comparison of the properties of BGPS
sources with and without an associated methanol maser shows that
those with an associated class I methanol maser usually have higher
beam-averaged H$_2$ column density and larger BGPS flux density than
those without an associated maser.

A series of investigations of the detection rates of class I
methanol masers as a function of GLIMPSE mid-IR and BGPS properties
were undertaken, with the aim of developing more efficient criteria
for future targeted class I methanol maser searches. Although the
detection rates of class I methanol masers appear to some extent to
be dependent on the mid-IR properties of GLIMPSE point sources (such
as 4.5 $\mu$m magnitude and [3.6]-[4.5] color), tighter correlations
are observed between the class I methanol maser detection rate and
the BGPS source properties. This suggests that the BGPS catalog
could provide much more efficient target samples for future class I
methanol maser searches. Based on the observed relationship between
the detection rate of class I methanol maser and the BGPS
beam-averaged H$_2$ column density, we estimate that approximately
1000 (of $\sim$8400) BGPS sources may have an associated class I methanol maser.
 We identify a region in the distribution of BGPS
beam-average column density versus BGPS integrated flux density
(satisfying $log(S_{int})\leq-38.0+1.72log(N_{H_{2}}^{beam})$, and
 $log(N_{H_{2}}^{beam})\geq 22.1$), towards which we we expect to find 90\%
of all ($\sim1000$) class I methanol masers with a high detection
efficiency ($\sim$75\%).

\acknowledgements

We thank an anonymous referee for their helpful comments which have
improved this paper. We are grateful to the staff of Qinghai Station
of Purple Mountain Observatory for their assistance in the
observation. This research has made use of the data products from
the GLIMPSE survey, which is a legacy science program of the {\em
Spitzer Space Telescope} funded by the National Aeronautics and
Space Administration, and made use of information from the BGPS
survey database at
http://irsa.ipac.caltech.edu/data/BOLOCAM$_-$GPS/. This work is
partly supported by China Ministry of Science and Technology under
State Key Development Program for Basic Research (2012CB821800), the
National Natural Science Foundation of China (grants 10621303,
10625314, 10803017, 10821302, 11073041, 11073054, 11133008,
11173046), the CAS/SAFEA International Partnership Program for
Creative Research Teams, and Key Laboratory for Radio Astronomy, CAS.
.


\tabletypesize{\tiny}

\setlength{\tabcolsep}{0.03in}

\begin{deluxetable}{llcccccclcccc}
\tablewidth{0pt} \tablecaption{Sample parameters}
\renewcommand{\arraystretch}{1.0}
\tablehead{ & \multicolumn{5}{c}{GLIMPSE Point Source} & & \multicolumn{6}{c}{BGPS Source} \\
\cline{2-6} \cline{8-13}
Number & \multicolumn{1}{c}{Name} & 3.6 $\mu$m & 4.5 $\mu$m & 5.8 $\mu$m  & 8.0 $\mu$m & & ID & \multicolumn{1}{c}{Name} & R & S$_{40''}$ & S$_{int}$ & $N_{H_{2}}^{beam}$\\
 & & (mag) & (mag) & (mag) & (mag) & & & & ($''$) & (Jy) & (Jy) & (10$^{22}$ cm$^{-2}$)\\
 (1) & \multicolumn{1}{c}{(2)} & (3) & (4) & (5) & (6) & & (7) & \multicolumn{1}{c}{(8)} & (9) & (10) & (11) & (12)}

\startdata

1    &   G004.7808$+$00.0429    &  12.12(0.04)  &  10.09(0.04)  &   8.96(0.03)  &    8.19(0.04)  &   &  1051   &   G004.783$+$00.043   &  27.96   &   0.175(0.041)  &  0.422(0.081)  & 1.05 \\
2    &   G004.8020$+$00.1880    &  13.30(0.06)  &  11.06(0.05)  &   9.99(0.04)  &    9.56(0.05)  &   &  1053   &   G004.803$+$00.191   &  25.94   &   0.094(0.038)  &  0.211(0.065)  & 0.57 \\
3    &   G004.9676$+$00.0500    &  13.00(0.05)  &  10.83(0.06)  &   9.07(0.03)  &    7.91(0.02)  &   &  1066   &   G004.967$+$00.047   &  --      &   0.096(0.041)  &  0.166(0.060)  & 0.58 \\
4    &   G005.0424$-$00.0977    &  13.31(0.07)  &  11.09(0.08)  &   9.57(0.04)  &    8.47(0.03)  &   &  1071   &   G005.043$-$00.099   &  57.77   &   0.157(0.041)  &  0.553(0.110)  & 0.95 \\
5    &   G005.3294$-$00.0949    &  14.04(0.11)  &  11.42(0.07)  &   9.81(0.04)  &    8.74(0.03)  &   &  1084   &   G005.333$-$00.093   &  --      &   0.087(0.039)  &  0.198(0.063)  & 0.52 \\
6    &   G005.3608$+$00.0179    &  12.39(0.08)  &  10.21(0.05)  &   8.76(0.03)  &    7.82(0.03)  &   &  1087   &   G005.361$+$00.017   &  46.14   &   0.281(0.041)  &  0.925(0.119)  & 1.69 \\
7    &   G005.3706$+$00.3179    &   9.55(0.09)  &   7.45(0.06)  &   6.00(0.02)  &    5.37(0.03)  &   &  1090   &   G005.373$+$00.319   &  --      &   0.145(0.036)  &  0.230(0.054)  & 0.87 \\
8    &   G005.8418$-$00.3756    &  13.34(0.09)  &  10.99(0.06)  &   9.17(0.04)  &    8.01(0.04)  &   &  1131   &   G005.841$-$00.379   &  --      &   0.107(0.058)  &  0.203(0.091)  & 0.64 \\
\enddata

\tablecomments{Column (1): source number which is organized by
increasing galactic longitude. Column (2): GLIMPSE point source
name. Columns (3) -- (6): the magnitude of the GLIMPSE point source
in the 3.6, 4.5, 5.8 and 8.0 $\mu$m bands, respectively. Columns (7)
and (8): the ID number and name of BGPS source, respectively. Column
(9): the radius of BGPS, sources which are unresolved with the BGPS
beam, are indicated with  ``--'' in this column. Columns (10) and
(11): the aperture flux density within 40$\arcsec$ and the
integrated flux density of the BGPS sources. Note that a flux
calibration correction factor of 1.5 should be applied to the both
the aperture and integrated flux densities listed here to calculate
BGPS gas mass and column/volume density (see Section 4.2). In
addition, an aperture correction of 1.46 is needed to apply to
aperture flux density within 40$\arcsec$ after applied a flux
calibration correction factor of 1.5 to calculate the beam-averaged
column density. Columns (12): the beam-averaged H$_{2}$ column
density (see section 4.2).}

\end{deluxetable}


\tabletypesize{\scriptsize}

\setlength{\tabcolsep}{0.07in}

\begin{deluxetable}{llccc|llccc}

\tablewidth{0pt} \tablecaption{Observed source positions and
observing rms noise.}

\tablehead{Number & BGPS ID & R.A. (J2000) & Decl. (J2000) &
$\sigma_{rms}$ (Jy)  & Number & BGPS ID & R.A. (J2000) & Decl.
(J2000) & $\sigma_{rms}$ (Jy)  }

\startdata

1    &   1051  &   17  56  25.86  &  -24  48  17.0  &   0.9    &   108  &   3530  &   18  38  35.11  &  -06  41  27.1  &   1.2   \\
2    &   1053  &   17  55  54.71  &  -24  42  46.7  &   0.9    &   109  &   3699  &   18  39  56.20  &  -05  38  48.3  &   0.9   \\
3    &   1066  &   17  56  49.37  &  -24  38  37.0  &   1.1    &   110  &   3721  &   18  38  57.67  &  -05  14  41.2  &   1.0   \\
4    &   1071  &   17  57  32.79  &  -24  39  03.9  &   0.9    &   111  &   3741  &   18  39  54.38  &  -05  10  05.3  &   0.4   \\
5    &   1084  &   17  58  09.70  &  -24  23  49.2  &   0.5    &   112  &   3771  &   18  40  39.47  &  -05  00  21.0  &   0.9   \\
6    &   1087  &   17  57  48.30  &  -24  19  03.8  &   0.9    &   113  &   3822  &   18  41  20.84  &  -04  32  20.6  &   1.0   \\
7    &   1090  &   17  56  41.11  &  -24  09  21.4  &   1.0    &   114  &   3833  &   18  42  56.88  &  -04  41  57.6  &   1.0   \\
8    &   1131  &   18  00  21.60  &  -24  05  56.6  &   0.9    &   115  &   3863  &   18  43  52.92  &  -04  36  19.3  &   1.1   \\
\enddata

\end{deluxetable}


\tabletypesize{\scriptsize}

\setlength{\tabcolsep}{0.1in}

\begin{deluxetable}{lrccccccccccc}

\tablewidth{0pt} \tablecaption{Observed properties of 95 GHz class I
methanol maser sources detected -- The full table is available in
the online journal.}
\tablehead{  &  \multicolumn{4}{c}{Gaussian Fit} & \\
\cline{2-5}
Number &  V$_{LSR}$ &  $\Delta V$  & $S$ & P & S$_{int}^m$ \\
 & (km s$^{-1}$) & (km s$^{-1}$) & (Jy km s$^{-1}$) & (Jy) & (Jy km s$^{-1}$)  \\
  (1) & (2) & (3) & (4) & (5) & (6) }

\startdata

11       &    20.36(0.08)   &        0.89(0.21)    &      5.8(1.1)    &     6.1   &  23.4    \\
...      &    21.35(0.07)   &        0.51(0.16)    &      3.2(1.0)    &     5.8   &           \\
...      &    22.63(0.06)   &        0.97(0.15)    &     10.4(1.5)    &     10.0   &           \\
...      &    24.02(0.18)   &        1.08(0.43)    &      4.1(1.4)    &     3.5   &           \\
13       &    20.08(0.89)   &       11.54(2.26)    &     15.0(2.5)    &     1.2   &  17.3    \\
...      &    20.62(0.12)   &        0.87(0.30)    &      2.3(0.8)    &     2.5   &           \\
20       &    98.48(0.30)   &        2.71(0.70)    &      6.9(1.5)    &     2.4   &  6.9     \\
22       &    -0.46(0.05)   &        0.26(0.22)    &      1.5(0.6)    &     5.4   &  20.9    \\
\enddata

\tablecomments{Column (1): source number. Columns. (2)-(5): the velocity
at peak V$_{LSR}$, the line FWHM $\Delta$V, the integrated
intensity $S$, and the peak flux density $P$ of each maser feature
estimated from Gaussian fits to the 95 GHz class I methanol maser lines.
The formal error from the Gaussian fit is given in parenthesis. Col.
(6): the total integrated flux density S$_{int}^m$ (Jy) of the maser spectrum
obtained from summing the integrated flux
density of all maser features in each source in column (4).}

\end{deluxetable}


\tabletypesize{\scriptsize}

\setlength{\tabcolsep}{0.1in}


\begin{deluxetable}{llccc}

\tablewidth{0pt} \tablecaption{Sources detected as class I methanol masers in previous
surveys.}

\tablehead{ Number & Source & \multicolumn{2}{c} {Detections} &
Referee \\
\cline{3-4}
 & & 44 GHz & 95 GHz & }

 \startdata

22     &     18024-2119   &  Y  &  N  &  Fontani et al. (2010) \\
32     &     G10.47+0.03  &  Y  & --  &  Kurtz et al. (2004) \\
33     &     G10.62-0.38  &  Y  & --  &  Kurtz et al. (2004) \\
43     &     EGO G12.20-0.03 &  & Y   &  Chen et al. (2011)  \\
51     &     EGO G12.91-0.03 &  & Y   &  Chen et al. (2011)  \\
69     &     EGO G16.59-0.05 & Y & Y   &  Slysh et al. (1994); Val'tts et al. (2000)  \\
73     &     EGO G18.89-0.47 & Y & Y   &  Chen et al. (2011)  \\
74     &     EGO G19.01-0.03 & Y & Y   &  Chen et al. (2011)  \\
76     &     EGO G19.88-0.53 & -- & Y  &  Chen et al. (2011)  \\
83     &     EGO G22.04+0.22 & Y  & Y  &  Chen et al. (2011)  \\
90     &     G23.43-0.19     & Y & --  &  Slysh et al. (1994) \\
93     &     EGO G23.96-0.11 & Y  & -- &  Chen et al. (2011)  \\
94     &     EGO G24.00-0.10 & -- & Y  &  Chen et al. (2011)  \\
97     &     EGO G24.33+0.14 & -- & Y  &  Chen et al. (2011)  \\
98     &     EGO G24.63+0.15 & -- & Y  &  Chen et al. (2011)  \\
101    &      W42            & Y  & -- &  Bachiller et al. (1990) \\
104    &     EGO G25.38-0.15 & -- & Y  &  Chen et al. (2011) \\
161    &     EGO G34.41+0.24 & -- & Y  &  Chen et al. (2011) \\
181    &     EGO G43.04-0.45 & Y  & -- &  Chen et al. (2011) \\
194    &     EGO G49.27-0.34 & Y  & -- &  Cyganowski et al. (2009) \\

\enddata
\end{deluxetable}


\tabletypesize{\scriptsize}

\setlength{\tabcolsep}{0.07in}


\begin{deluxetable}{llcccccccc}

\tablewidth{0pt} \tablecaption{The related parameters of the
detected class I methanol masers.}

\tablehead{ Number & Distance & Luminosity & \multicolumn{4}{c}{BGPS source} & IRDC & Class II maser \\
 \cline{4-7}
 & & & ID  & M & n(H$_{2}$) & N(H$_{2}$) & \\
 & (kpc) & ($10^{-6} L_\odot$) &  & (M$_{\odot}$) & (10$^3$ cm$^{-3}$) & (10$^{22}$ cm$^{-2}$)
 \\
 (1) & (2) & (3) & (4) & (5) & (6) & (7) & (8) & (9)}

 \startdata

11                  &   13.0   &  124.3  &  1251 &    4900       &   1.6    &    1.5    &  N  &   N  \\
13                  &    3.4   &    6.3  &  1289 &     310       &   5.3    &    1.4    &  Y  &   N  \\
20$^{D}$            &    4.3   &    4.0  &  1363 &     370       &   4.8    &    1.4    &  Y  &   N  \\
22$^{G,D}$          &    5.2   &   17.8  &  1395 &    2100       &   1.5    &    1.1    &  Y  &   Y  \\
29$^{S}$            &    1.9   &    0.9  &  1467 &     560       &   3.0    &    1.1    &  Y  &   N  \\
31                  &   12.7   &  229.6  &  1479 &   11000       &   0.8    &    1.2    &  N  &   Y  \\
32                  &   11.2   &  308.6  &  1497 &   52000       &  24.6    &   20.8    &  N  &   Y  \\
33                  &    4.0$^\ast$   &   57.0  &  1508 &    6300       & 113.9    &   28.7    &  Y  &   Y  \\
\enddata

\tablecomments{Column (1): source number. The sources which are
marked by $G$, $D$ or $S$ overlaid with that in Green \&
McClure-Griffiths (2011), Dunham et al. (2011b) or Schlingman et al.
(2011), respectively. Column (2): the kinematic distance for the
source, estimated from the Galactic rotation curve of Reid et al.
(2009). For sources overlapped with Green \& McClure-Griffiths
(2011), we adpoted the distances estimated from their work. For the
sources (marked by $\ast$) of which distances cannot be derived from
the Galactic rotation curve, a distance of 4 kpc is adopted for
source with an IRDC association (N33), and that determined in
Schlingman et al. (2011) for the other two sources (N194 and N210).
Column (3): the integrated luminosity of 95 GHz methanol maser.
Columns (4) -- (7): the ID number of BGPS source in the BGPS
catalog, the derived gas mass and averaged H$_{2}$ volume and column
densities of the BGPS source, respectively. The gas volume and
column densities can not be determined due to absence of radius
information for the sources which are unresolved by the BGPS beam,
we marked them with ``--''. Column (8): association with IRDC: Y =
Yes, N = No. Column (9): association with a 6.7 GHz methanol maser
for which a precise position has been measured. The positions of the
6.7 GHz class II methanol masers were identified from published 6.7
GHz maser catalogs (Caswell 2009; Caswell et al. 2010; Green et al.
2010; Caswell et al. 2011 ; Green et al. 2012): Y = Yes, N =
No,``--'' = no information.}

\end{deluxetable}


\begin{deluxetable}{lcccccccc}
\tabletypesize{\scriptsize}

\tablewidth{0pt} \tablecaption{Trends with star formation activity
for sources with and without methanol masers}

\tablehead{ Property & Group & mean & standard deviation & minimum &
median & maximum }

\startdata
\multicolumn{7}{c}{With and without class I methanol masers} \\

log($N_{H_{2}}^{beam}$)  & class I &      22.7  &   0.4  & 21.9 &  22.7  &  23.8 \\
  $[cm^{-2}]$            & no class I &   21.9  &   0.3  & 21.4 &  21.9  &  22.7 \\
\\
log($S_{int}$)           & class I &       0.7  &   0.4  & -0.6 &  0.7   &  1.5  \\
  $[Jy]$                 & no class I &    0.0  &   0.4  & -0.9 &  0.0   &  1.1  \\
\\
Radius                   & class I &     57.8  &  25. 1  & 15.7 &  56.0  &  128.7 \\
$('')$                   & no class I &  49.3  &  21.4   &  5.1 &  46.3  &  113.3   \\

\hline
\multicolumn{7}{c}{Class I methanol masers with and without class II maser associations}\\

log($N_{H_{2}}^{beam}$)  & only class I & 22.5  &  0.2  &  22.2  &  22.6 & 22.8  \\
  $[cm^{-2}]$            &  class I+II  & 22.9  &  0.3  &  22.5  &  22.8 & 23.8  \\
  \\
log(S$_{int}$)           & only class I & 0.48  &  0.29 &  -0.09 &  0.47 & 1.07  \\
  $[Jy]$                 &  class I+II  & 0.86  &  0.31 &   0.41 &  0.79 & 1.49  \\

\enddata

\end{deluxetable}


\begin{deluxetable}{lcccccccc}
\tabletypesize{\scriptsize} \setlength{\tabcolsep}{0.05in}

\tablewidth{0pt} \tablecaption{The distributions of class I methanol
maser numbers with BGPS source parameters}

\tablehead{ Range & \multicolumn{3}{c}{In our observing sample} & &
\multicolumn{2}{c}{In the full BGPS catalog}\\
\cline{2-4} \cline{6-7}
 & detections & total numbers & detection rate & & total numbers &
 expected detections}

\startdata
log($N_{H_{2}}^{beam}$) $[cm^{-2}]$ \\

21.0--21.2  &  -- & -- & --    & &   1    &  0       \\
21.2--21.4  &  -- & -- & --    & &   88   &  0       \\
21.4--21.6  &  0  & 1  & 0.00  & &  717   &  0       \\
21.6--21.8  &  0  & 11 & 0.00  & &  2021  &  0       \\
21.8--22.0  &  1  & 46 & 0.02  & &  2303  &  50      \\
22.0--22.2  &  6  & 45 & 0.13  & &  1451  &  193     \\
22.2--22.4  &  6  & 36 & 0.17  & &  804   &  134     \\
\enddata

\end{deluxetable}

\clearpage

\begin{figure*}
\scalebox{0.50}[0.50]{\includegraphics[-80,0][600,420]{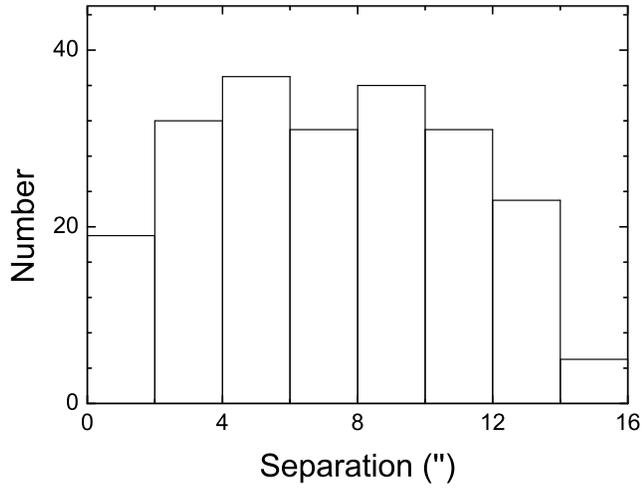}}
\caption{Number of sources as a function of the separations of the
pair of GLIMPSE point source and GBPS source in our observing
sample.}
\end{figure*}

\begin{figure*}
\scalebox{1.1}[1.1]{\includegraphics[90,10][500,400]{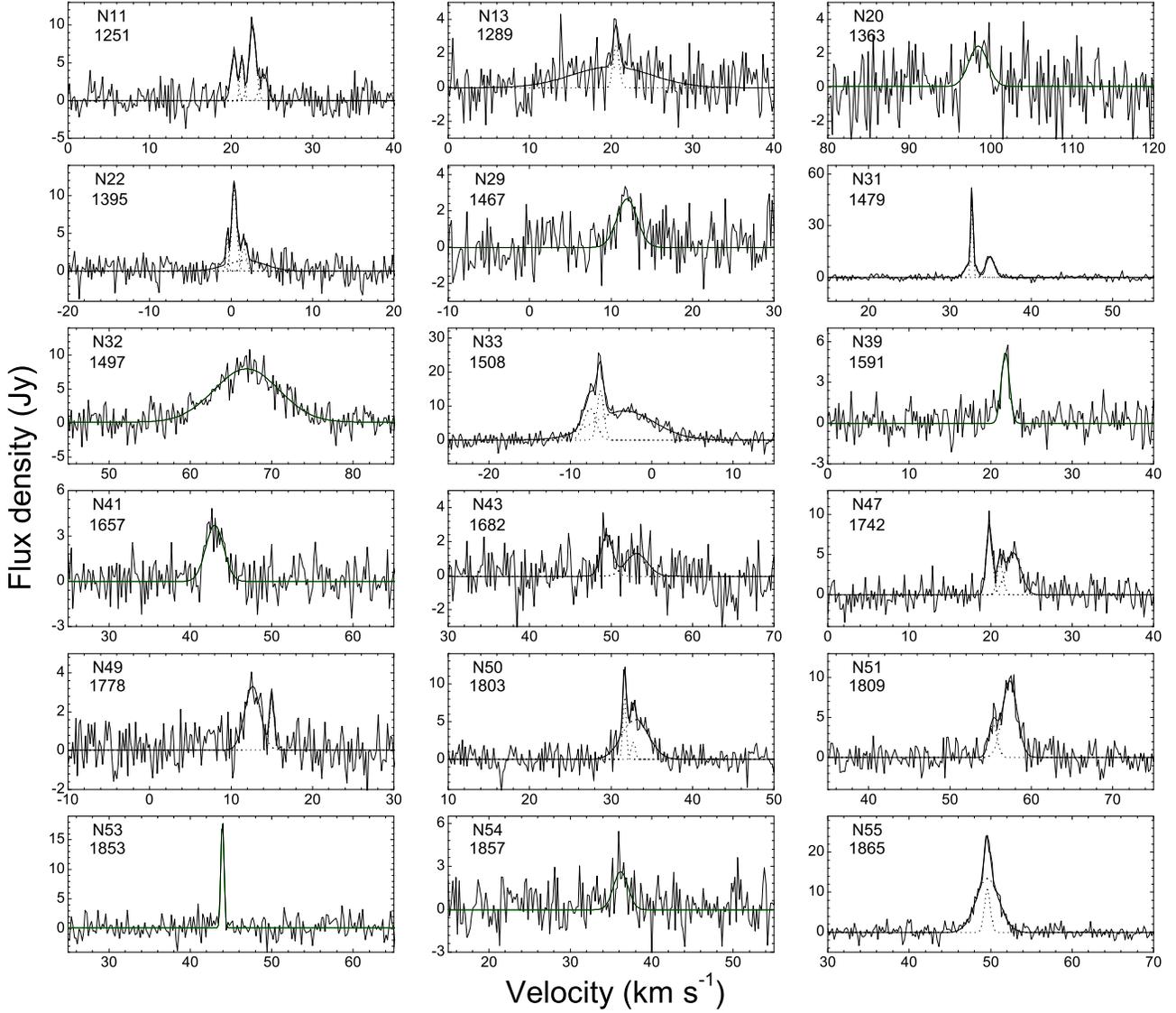}}
\caption{Spectra of the 95 GHz methanol masers detected in the
survey. The dashed lines represent the Gaussian fitting of each
maser feature, the bold-solid line mark the sum of the Gaussian
fitting of all maser feature.}

\end{figure*}

\begin{figure*}
\scalebox{1.1}[1.1]{\includegraphics[90,10][500,400]{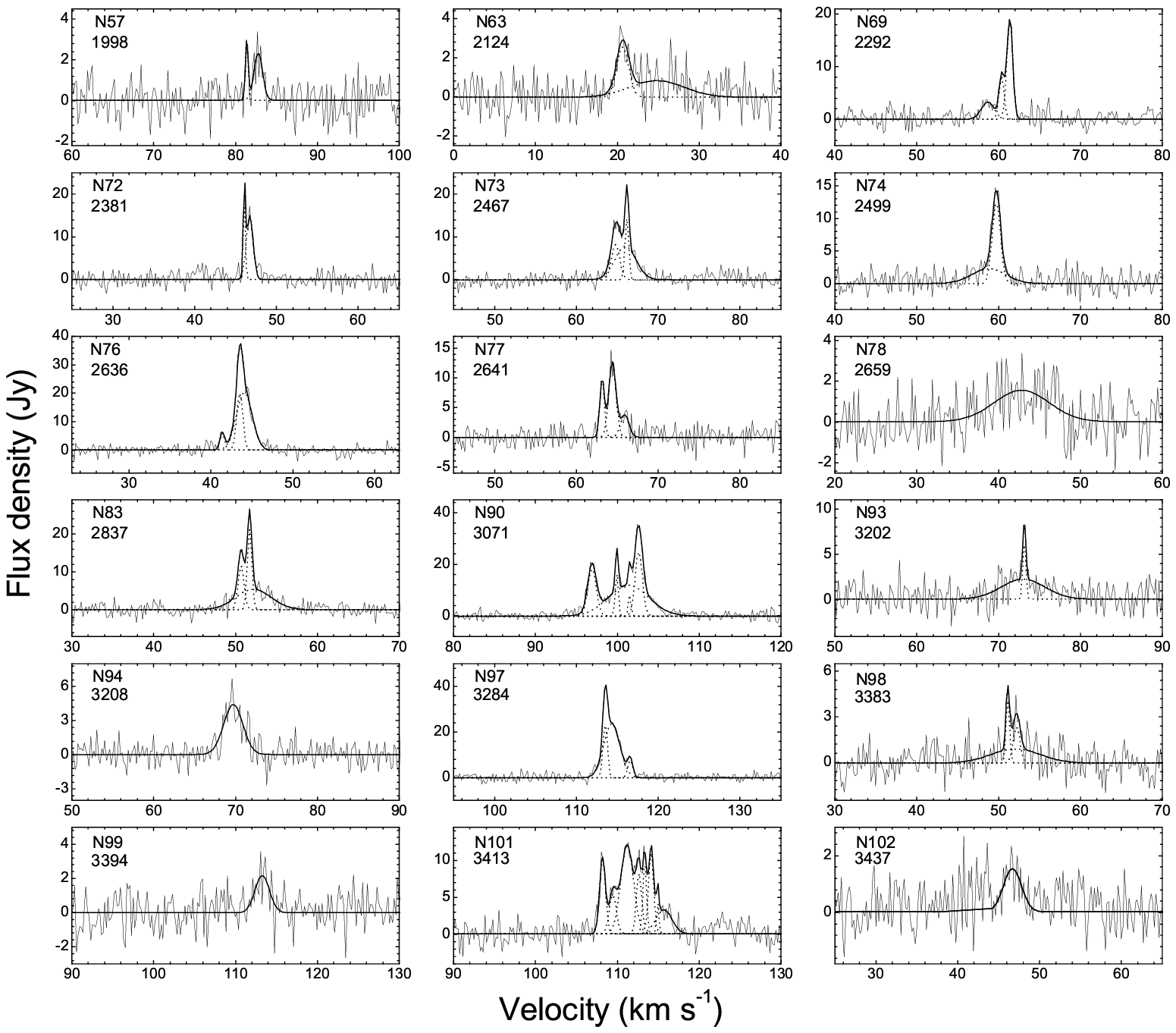}}
\vspace{6mm}

Fig. 2.--- Continued.
\end{figure*}

\begin{figure*}
\scalebox{1.1}[1.1]{\includegraphics[90,10][500,400]{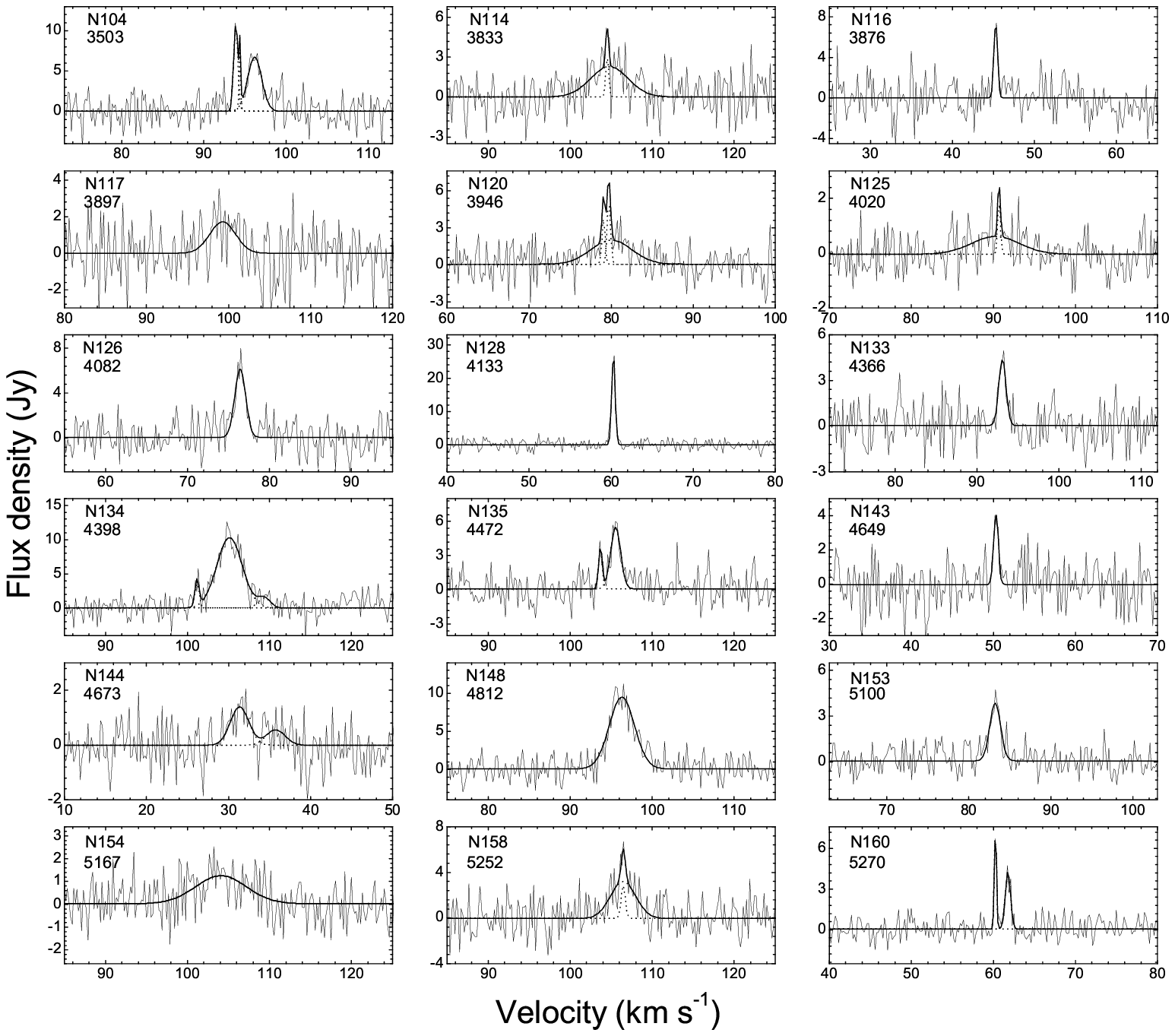}}
\vspace{6mm}

Fig. 2.--- Continued.
\end{figure*}

\begin{figure*}
\scalebox{1.1}[1.1]{\includegraphics[90,200][500,400]{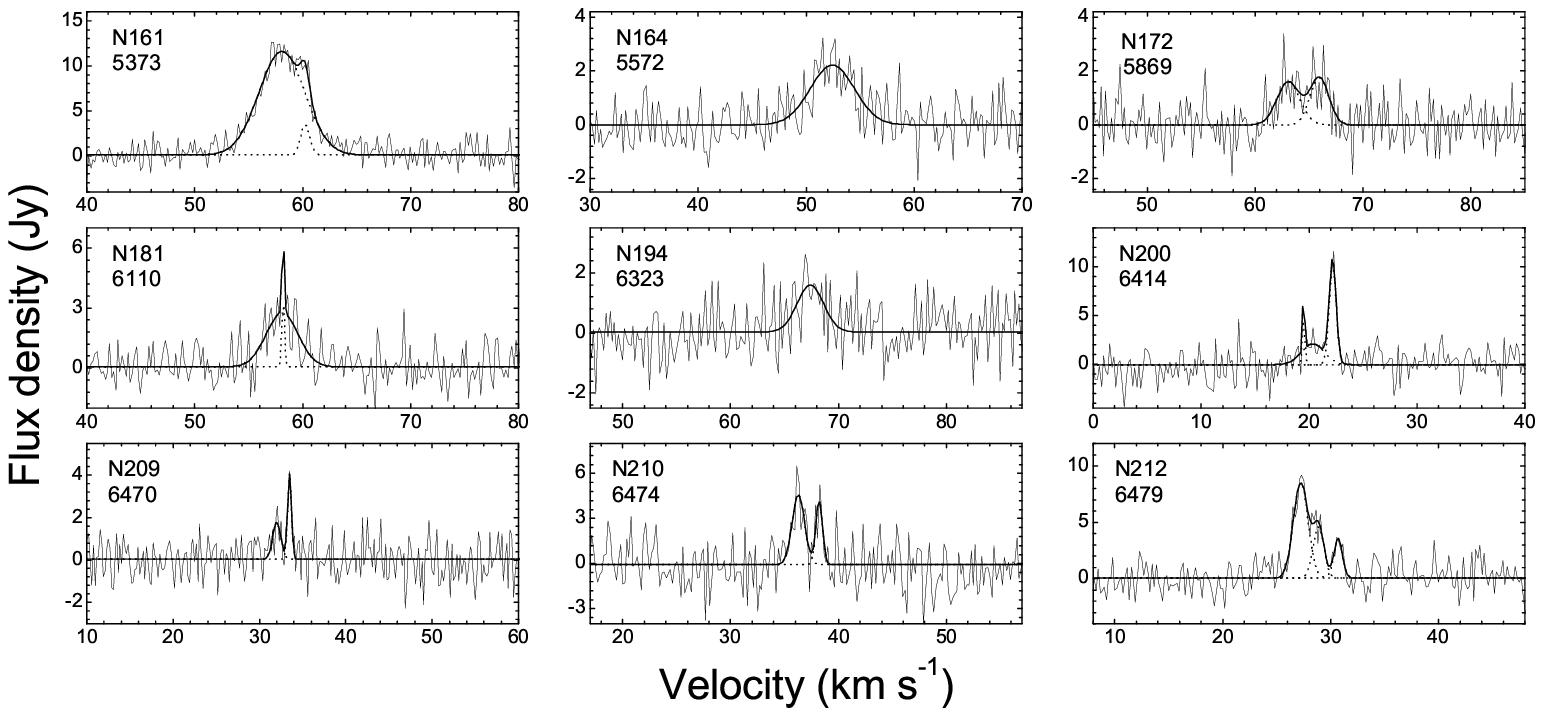}}
\vspace{6mm}

Fig. 2.--- Continued.
\end{figure*}

\begin{figure*}
\scalebox{1.1}[1.1]{\includegraphics[90,40][600,400]{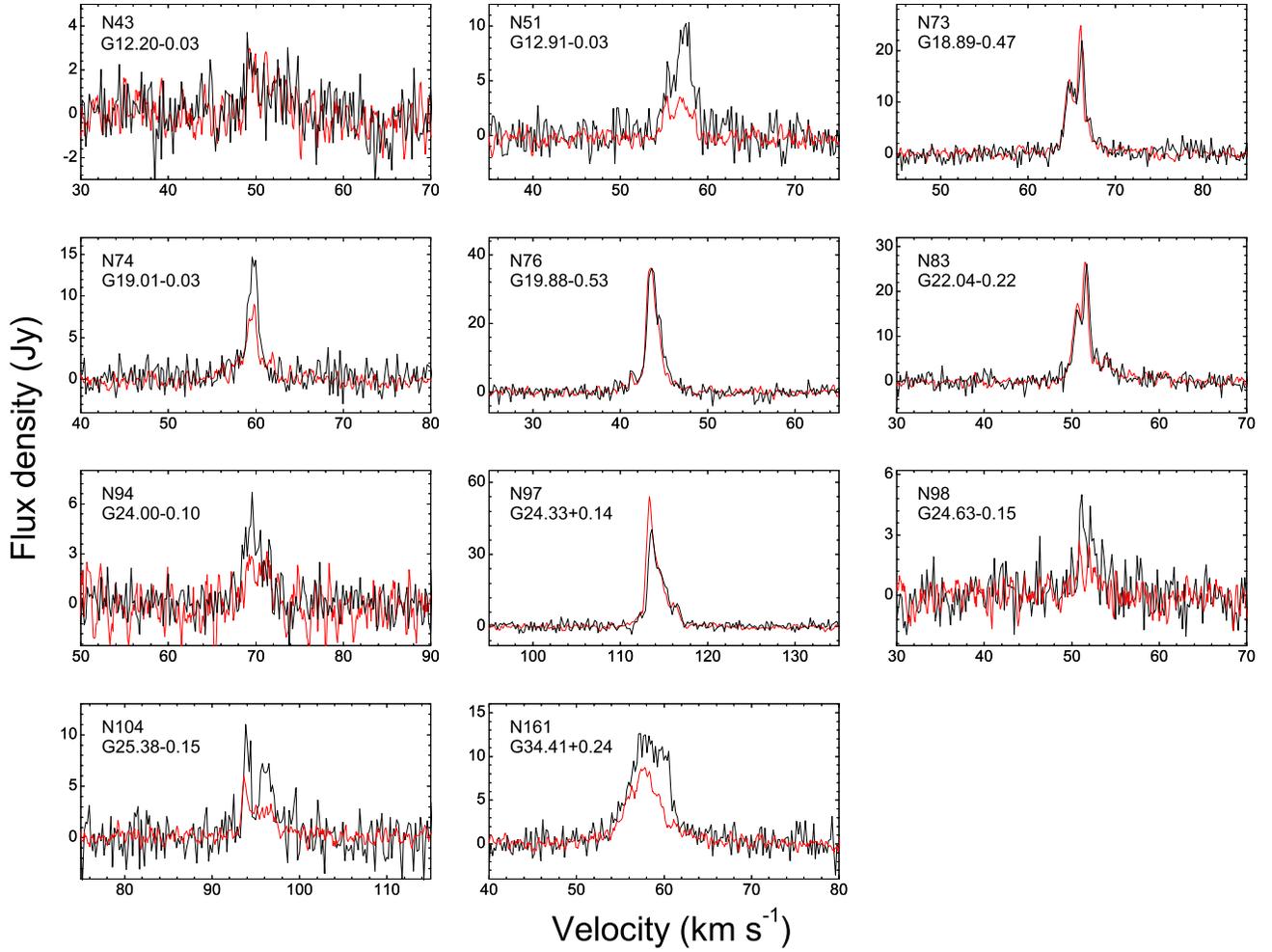}}
\caption{Comparison of the spectra of 95 GHz methanol maser emission
in the 11 sources which have been detected in both the PMO 13.7-m
survey (this work) marked with black lines and the EGO-based Mopra
survey by Chen et al. (2011) marked with red lines. A color version
of this figure is available in the online journal.}
\end{figure*}

\begin{figure*}
\scalebox{0.9}[0.9]{\includegraphics[30,190][600,400]{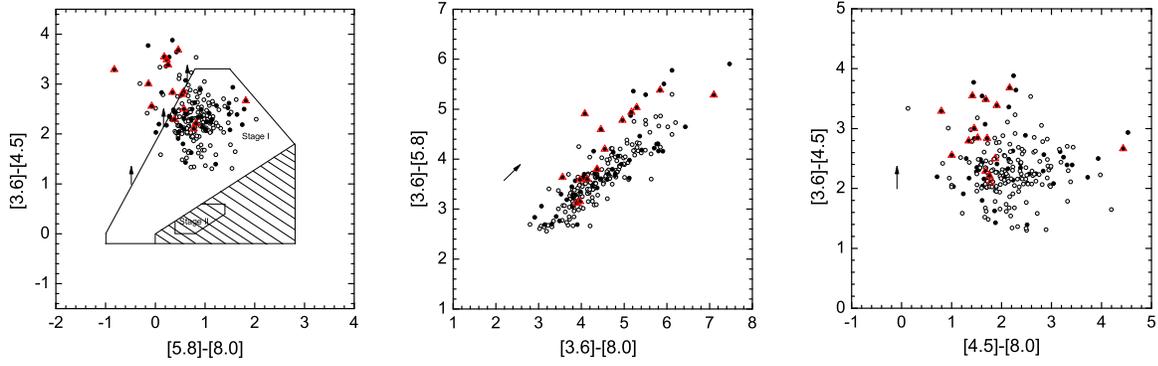}}
\caption{Color-color diagrams of GLIMPSE point sources associated
with and without class I methanol maser detections in the survey.
Filled and open circles represent the sources with and without class
I methanol maser detections, respectively. The sources associated
with EGOs (15 in total) are enclosed by red triangles. The solid
lines overlaid in [3.6]-[4.5] vs. [5.8]-[8.0] diagram construct the
regions occupied by various evolutionary-stage (Stages I, II and
III) YSOs according to the models of Robitaille et al. (2006). The
hatched region in the color-color plot is the region where models of
all evolutionary stages can be present. Note that the Stage II area
in the color-color plot is hatched to show that most models in this
region are Stage II models, however Stage I models can also be found
within this area. The reddening vectors in each panel show an
extinction of A$_{v}$=20, assuming the Indebetouw et al. (2005)
extinction law. A color version of this figure is available in the
online journal. }
\end{figure*}

\begin{figure*}
\scalebox{0.8}[0.8]{\includegraphics[20,130][600,400]{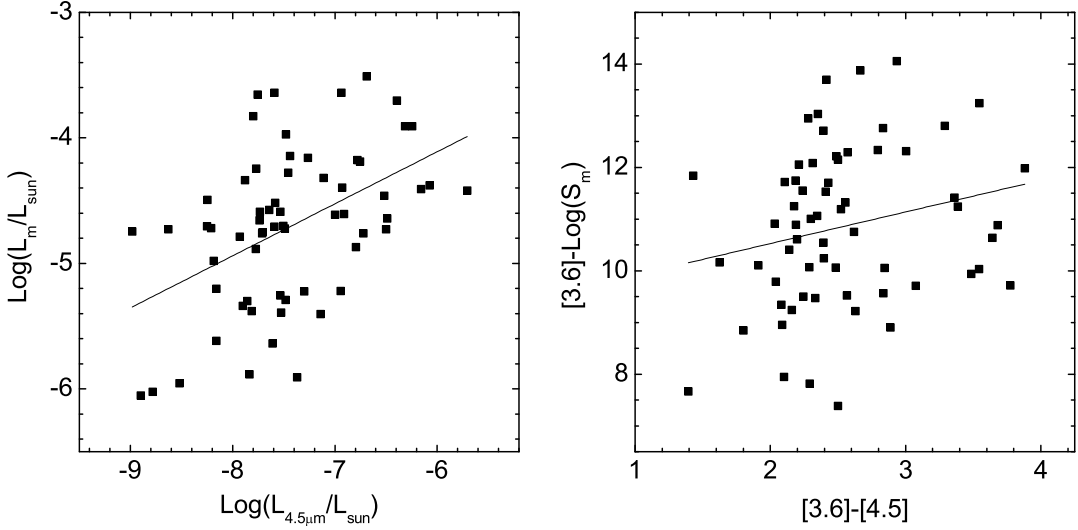}}
\caption{Left: logarithm of the 95 GHz class I methanol maser
luminosity versus GLIMPSE point source luminosity at 4.5 $\mu$m
band; Right: color-color diagram of [3.6]-log(S$_{m}$) versus
[3.6]-[4.5] which combines the GLIMPSE point sources and class I
methanol maser emission. The line in each panel marks the best fit
to the corresponding distribution.}
\end{figure*}

\begin{figure*}
\scalebox{0.8}[0.8]{\includegraphics[20,130][600,400]{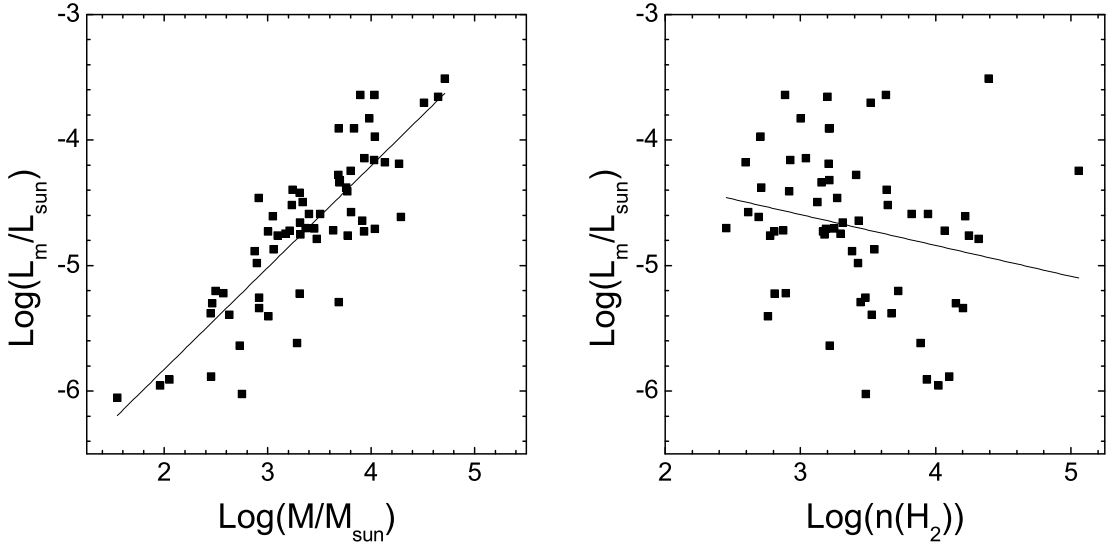}}
\caption{Logarithm of the 95 GHz class I methanol maser luminosity
as a function of the gas mass (left panel) and H$_{2}$ volume
density (right panel) of the associated 1.1 mm BGPS sources. The
line in each panel marks the best fit from the linear regression
analysis to the corresponding distribution. }
\end{figure*}

\begin{figure*}
\scalebox{0.6}[0.6]{\includegraphics[-80,0][600,400]{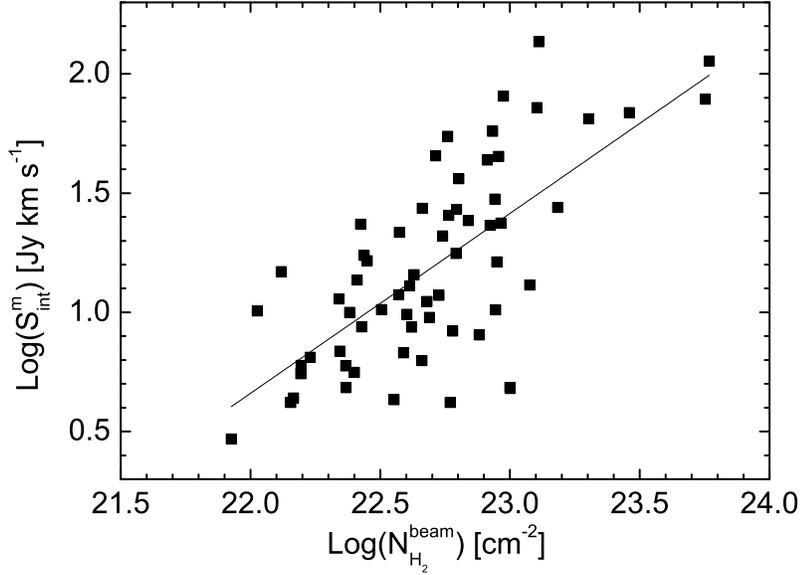}}
\caption{Logarithm of the integrated flux density of the 95 GHz
class I methanol maser as a function of the beam-averaged H$_2$
column density of the BGPS source. The line marks the best fit from
the linear regression analysis to the distribution. }
\end{figure*}

\begin{figure*}
\scalebox{0.8}[0.8]{\includegraphics[15,180][600,400]{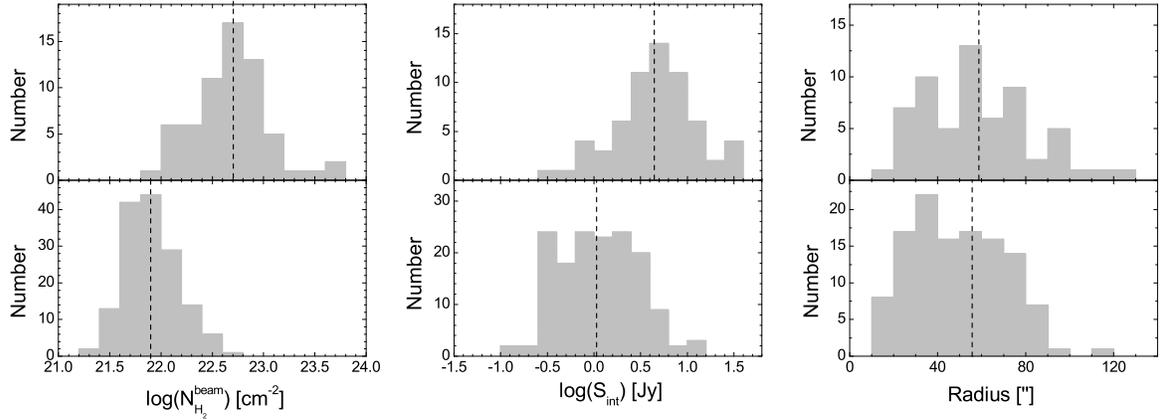}}
\caption{Number of sources as functions of the BGPS beam-averaged
H$_{2}$ column density (left), integrated flux density of the BGPS
source (middle) and BGPS source radius (right) for the two groups
with and without class I methanol maser detections. For
distributions in each BGPS property, the upper and lower panels
correspond to the BGPS sources with and without class I methanol
maser detections. The mean of each distribution is marked by the
vertical dashed line in the corresponding distribution.}
\end{figure*}

\begin{figure*}
\scalebox{0.8}[0.8]{\includegraphics[0,80][600,400]{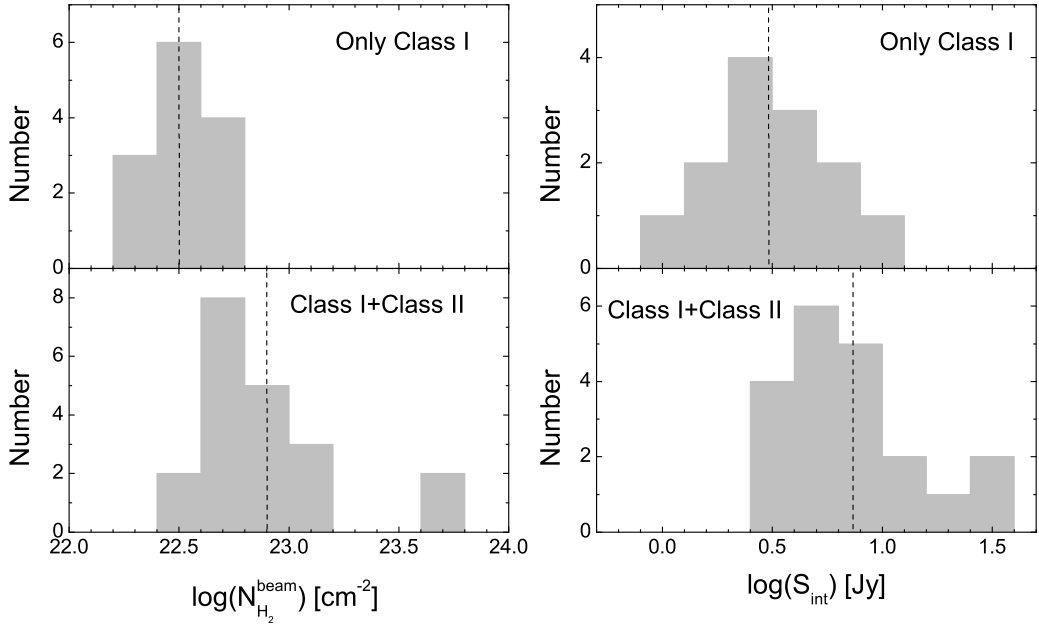}}
\caption{Number of sources as functions of the BGPS beam-averaged
H$_{2}$ column density (left) and BGPS integrated flux density
(right) for the two subsamples based on which class of methanol
masers they are associated with. For distributions in each BGPS
property, the upper and lower panels correspond to the BGPS sources
associated with only class I methanol masers and associated with
both class I and II methanol masers, respectively. The mean of each
distribution is marked by the vertical dashed line in the
corresponding distribution.}
\end{figure*}

\begin{figure*}
\scalebox{0.6}[0.6]{\includegraphics[-80,80][600,400]{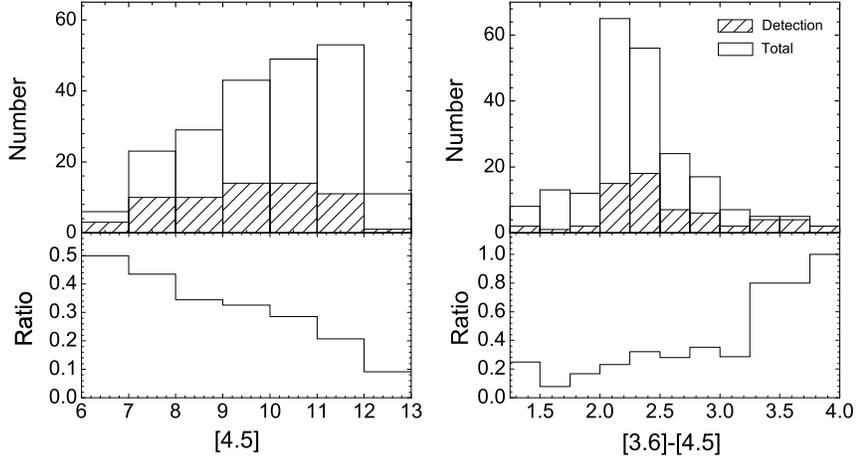}}
\caption{Detection rates of class I methanol masers with 4.5 $\mu$m
magnitude (left panel) and [3.6]-[4.5] color (right panel) of the
GLIMPSE point sources. For each mid-IR property, the upper panel
shows the histogram distributions of number of total sample sources
and detected class I methanol maser sources marked with open bars
and diagonal bars, respectively, and the lower panel shows the
corresponding detection rate of class I methanol maser in each
statistical bin.}
\end{figure*}

\begin{figure*}
\scalebox{0.8}[0.8]{\includegraphics[-80,170][600,400]{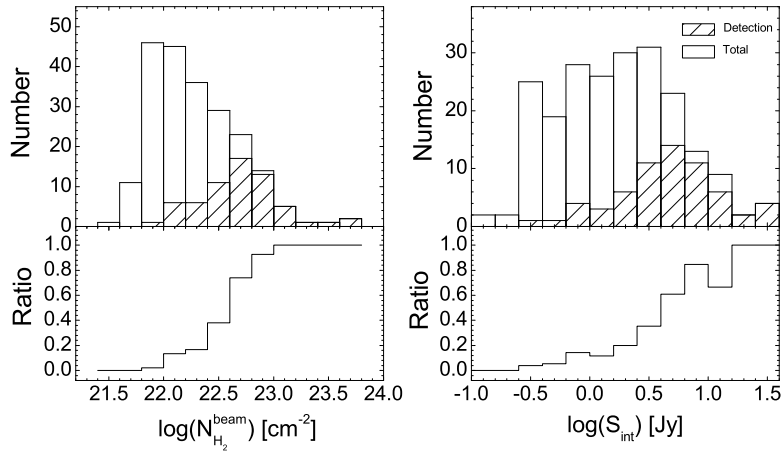}}
\caption{As Figure 10, but for detection rates of class I methanol
maser with the BGPS properties of the beam-averaged H$_{2}$ column
density (left), integrated flux density (right).}
\end{figure*}

\begin{figure*}
\scalebox{0.8}[0.8]{\includegraphics[0,130][600,400]{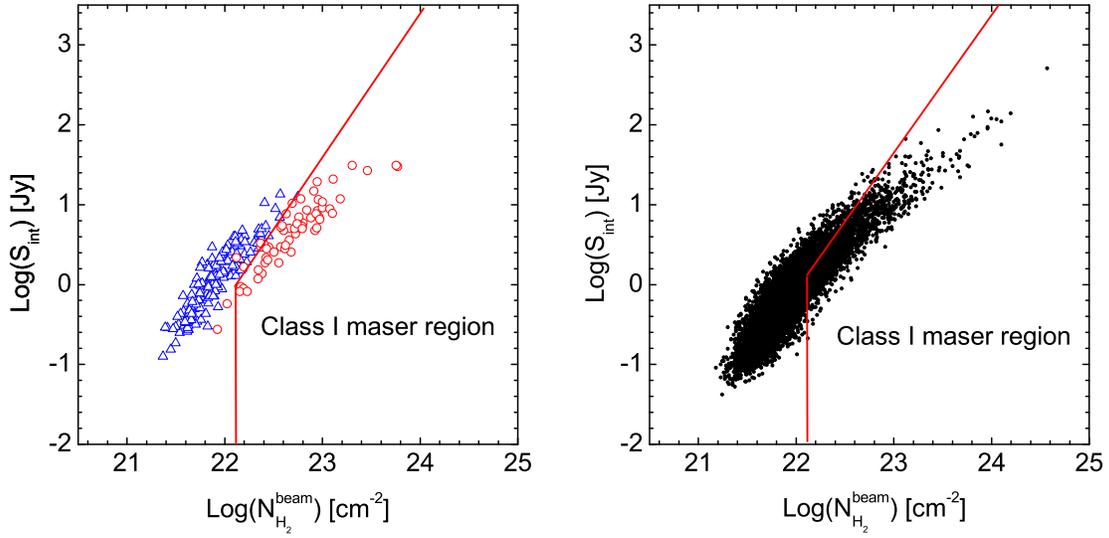}}
\caption{Left panel: Logarithm of the integrated flux densities
versus beam-averaged H$_{2}$ column density of BGPS sources with and
without class I methanol maser detections (marked by red circles and
blue triangles, respectively) in our current survey sample, the
class I methanol maser locating region is enclosed by the red lines.
Right panel: As Left panel, but for all cataloged BGPS sources. (A
color version of this figure is available in the online journal.)}

\end{figure*}

\end{document}